\documentclass[pdflatex,sn-basic,Numbered,iicol]{sn-jnl}

\usepackage{graphicx}\usepackage{multirow}\usepackage{amsmath,amssymb,amsfonts}\usepackage{amsthm}\usepackage{mathrsfs}\usepackage[title]{appendix}\usepackage{xcolor}\usepackage{textcomp}\usepackage{manyfoot}\usepackage{booktabs}\usepackage{listings}

\usepackage{pifont}\usepackage{tabularx}\usepackage{colortbl}\usepackage{makecell}
\usepackage[ruled,linesnumbered,vlined]{algorithm2e}
\usepackage[shortlabels]{enumitem} \usepackage{listings}\usepackage{subcaption}\usepackage{fdf}
\usepackage{placeins}
\usepackage[autostyle=false, style=english]{csquotes}
\MakeOuterQuote{"}
\graphicspath{{images/}}
\usepackage{xurl}

\SetAlgorithmName{Procedure}{Procedure}{List of Procedures}
\SetKwProg{Fn}{def}{:}{end}
\DontPrintSemicolon

\definecolor{darkred}{HTML}{8B0000}

\theoremstyle{thmstyleone}

\theoremstyle{thmstyletwo}

\theoremstyle{thmstylethree}

\newcommand{\footurl}[1]{\footnote{\url{#1}}}

\begin{document}

\newcommand{\Builder}{\textsc{DesCartes Builder}\xspace}
\title[Building real-time digital twin instances with Function+Data Flow]{Building real-time digital twin instances with Function+Data Flow: user evaluation and extension for iterative pipelines}

\author*[1,2]{\fnm{Eduardo} \sur{de Conto}} \email{eduardo002@ntu.edu.sg}
\author[2,3]{\fnm{Blaise} \sur{Genest}}\email{blaise.genest@cnrs.fr}
\author[1]{\fnm{Arvind} \sur{Easwaran}}\email{arvinde@ntu.edu.sg}
\author[2]{\fnm{Nicholas} \sur{Ng}}
\author[2]{\fnm{Shweta} \sur{Menon}}

\affil*[1]{
    \orgdiv{College of Computing and Data Science},
    \orgname{Nanyang Technological University},\orgaddress{
        \street{50 Nanyang Avenue}, \city{Singapore},
        \postcode{639798}, \country{Singapore}}}
\affil[2]{
    \orgname{CNRS@CREATE},\orgaddress{
        \street{1 CREATE Way}, \city{Singapore}, \postcode{138602}, \country{Singapore}}}
\affil[3]{\orgname{CNRS, IPAL Lab}, \country{France}}

\abstract{
    Digital twins (DTs) increasingly leverage artificial intelligence (AI) and machine learning (ML) pipelines, both to build real-time DTs from high-fidelity simulations and to instantiate them with historical data. However, engineering these pipelines remains largely ad-hoc: pipelines are hard to specify, validate, and reuse, with scarce dedicated tooling. 
Function+Data Flow (FDF) addresses this by defining a visual domain-specific language (DSL) that represents functions (ML models) explicitly, enabling their composition and reuse. We implemented FDF in {\Builder}, an integrated modeling environment supporting FDF-based DT synthesis and validation. 

In this paper, we report on an empirical user study evaluating whether FDF and {\Builder} can make AI-based DT development more accessible and reliable. Participants implemented a representative real-time DT prototype within {\Builder}, and we measured perceived usability and feature adequacy through quantitative and qualitative measures.
Our results indicate that {\Builder} and FDF achieve a \emph{good} level of usability across a broad range of potential users, and particularly for the intended audience of domain experts. The study additionally surfaces concrete strengths and areas for improvement of both the tool and the underlying FDF framework.
Informed by these findings, we propose {\em H-FDF}, a {\em Hierarchical} extension of FDF supporting iterative and modular pipelines, enabling the formal specification of more complex DT pipelines such as dual training. Our findings suggest that integrated, model-driven platforms are a promising direction to transform AI-based DT engineering into a disciplined modeling practice.

 }
\maketitle

\clearpage 

\section{Introduction} \label{sec:introduction}

A \emph{digital twin (DT)} is a software entity that accurately mirrors and co-evolves with its associated physical system, known as the \emph{physical twin (PT)}. DTs are increasingly employed in safety-critical domains, including smart grids and cities~\cite{wangShortTermWindSpeed2023,jafariReviewDigitalTwin2023,danilczykANGELIntelligentDigital2019},
manufacturing~\cite{moyaDigitalTwinsThat2022,ghnatiosHybridTwinBased2024,kritzingerDigitalTwinManufacturing2018}, and aviation~\cite{tuegelReengineeringAircraftStructural2011,utzigAugmentedRealityRemote2019,xiongDigitalTwinApplications2022}. In these fields, DTs are essential for failure prediction and continuous performance optimization. The DT market is expanding rapidly and is projected to grow to US\$ 180--250 billion by 2032 from approximately US\$ 13 billion today~\cite{DigitalTwinMarket,EmergingTechnologiesRevenue}.

\begin{figure}[t]
    \centering
    \includegraphics[width=0.95\linewidth]{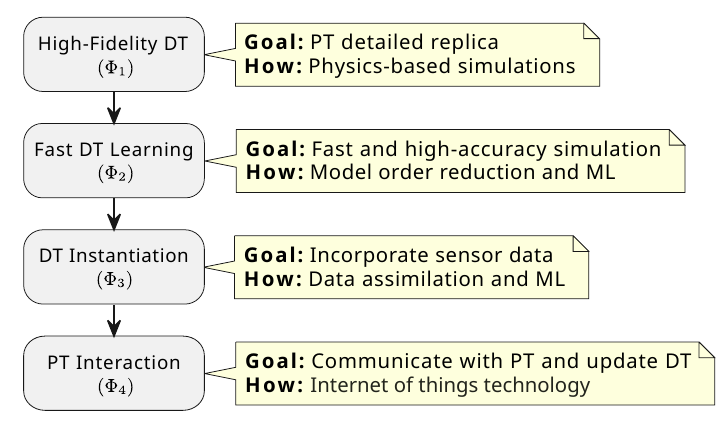}
    \caption{Summary of DT engineering phases. FDF~\cite{decontoFunction+DataFlowFramework2024} and {\Builder}~\cite{decontoDesCartesBuilderTool2025} focus on the second and third phases of fast DT learning~\ref{dt-phi-2} and DT instantiation~\ref{dt-phi-3}}
    \label{fig:dt-phases}
\end{figure}

However, despite the significance of DT and the growing market, its engineering remains a complex endeavor~\cite{tadejaMappingDigitalTwin2025}. This process can be structured into \emph{four phases}~\cite{grievesDigitalTwinMitigating2017}, denoted as~\ref{dt-phi-1}--\ref{dt-phi-4}, as depicted in Fig.~\ref{fig:dt-phases}: 
\begin{enumerate}[label={($\Phi_\arabic*$)}]
    \item\label{dt-phi-1} \emph{High-fidelity DT creation.} A \emph{DT prototype (DTP)} is created, potentially before the PT exists. This phase relies on high-fidelity physics-based simulations~\cite{glaessgenDigitalTwinParadigm2012,daliborCrossDomainSystematicMapping2022}, which enable accurate predictions across a wide range of parameters. However, as complexity increases, these simulations can become \emph{slow, computationally expensive} and numerically unstable~\cite{liuReviewDigitalTwin2021}: each new simulation can require hours or days, hindering DT use.
    
    \item\label{dt-phi-2} \emph{Fast DT learning}. 
    A \emph{real-time DTP} is created as a faster/real-time version of the DTP~\cite{mcclellanPhysicsbasedDigitalTwin2022}. This is typically achieved using \emph{model order reduction (MOR)} techniques, which enable significantly faster simulations without sacrificing accuracy~\cite{sancarlosLearningStableReducedorder2021,hartmannModelOrderReduction2018,chinestaShortReviewModel2011}. Hence, each DT simulation can now take on the order of seconds, a speed-up of about four orders of magnitude. MOR is often achieved using machine learning (ML), in particular \emph{unsupervised learning} techniques such as principal component analysis (PCA). 
    
    \item\label{dt-phi-3} \emph{DT instantiation.} The real-time DT prototype is \emph{specialized} to represent a specific PT instance, resulting in a real-time \emph{DT instance (DTI)}. This involves \emph{data assimilation}, in which historical sensor data from the PT is integrated with high-fidelity or reduced-order models (ROMs) derived in the earlier phases~\cite{donatoSelfupdatingDigitalTwin2024,zhongReducedorderDigitalTwin2023}. ML techniques, particularly supervised learning (e.g., neural networks trained via gradient descent) using sensor data as supervised input, are commonly employed. 

    \item\label{dt-phi-4} \emph{PT interaction.} The DT instance is \emph{continuously updated}, creating a closed-loop interaction with the PT. This involves acquiring live sensor data, interacting with the DT, and using the resulting information to control and monitor the corresponding physical system. This last phase is often achieved using \emph{internet of things (IoT)} technologies. \end{enumerate}

The outcome of each phase must be thoroughly validated. In~\ref{dt-phi-2}, the validation consists of comparing the real-time DTP predictions with the high-fidelity simulations from~\ref{dt-phi-1}. On the other hand, the real-time DTI~\ref{dt-phi-3} can be validated with historical data from the actual PT.
Note that DT models resulting from~\ref{dt-phi-2} and~\ref{dt-phi-3} are also referred to as \emph{surrogate} models~\cite{chakrabortyRoleSurrogateModels2021}. While the DT prototype~\ref{dt-phi-1} relies on simulations generated offline, these surrogate models can be evaluated in real-time during online deployment. In addition, these four phases are highly interdependent, often deviating from a linear workflow.

\begin{figure}[t!]
    \centering
    \includegraphics[width=\linewidth]{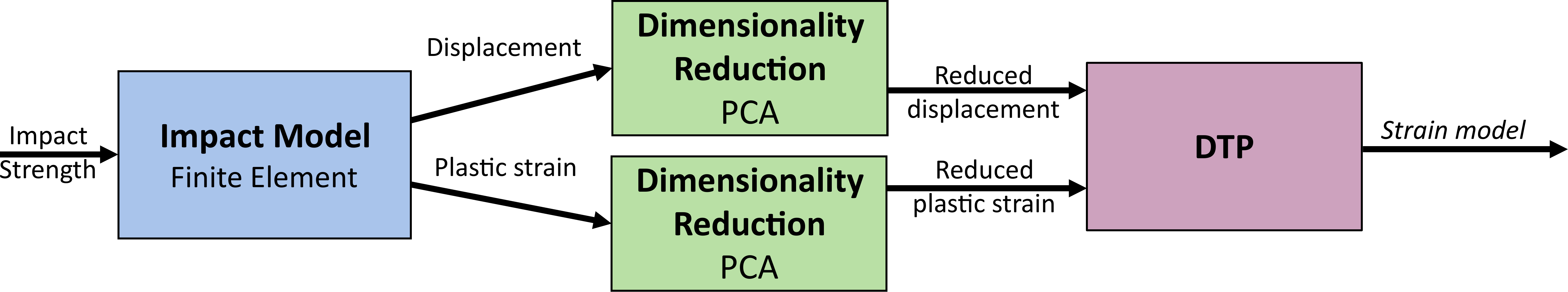}
    \caption{Pipeline to design a real-time DT prototype for material strain prediction (extracted from~\cite{decontoFunction+DataFlowFramework2024})}\label{fig:pipe-strain-overview}
\end{figure}

We now illustrate these different phases and their interaction on \emph{material strain prediction}, from the civil engineering domain. This is a typical use case of DTP that provides a predictive maintenance~\cite{tuegelReengineeringAircraftStructural2011} service via structural health monitoring~\cite{chabodDigitalTwinFatigue2022}. The \emph{goal} is to predict the plastic strain of a certain structure (which cannot be measured non-destructively) given an observed deformation, estimated using high-resolution 3D photos~\cite{johnsonComplexitiesCapturingLarge2023,tehraniPipeProfilingUsing2020}.
To design a real-time DTP using phases~\ref{dt-phi-1} and ~\ref{dt-phi-2}, the following pipeline, illustrated in Fig.~\ref{fig:pipe-strain-overview}, could be used:
\begin{itemize}
    \item[~\ref{dt-phi-1}] \emph{High-fidelity DT.} Use a (slow) finite element model (FEM) of the structure to accurately estimate the plastic strain and the deformation given different material parameters (impact strength, material thickness, etc.).
    \item[~\ref{dt-phi-2}] \emph{Model order reduction.} First, reduce the dimensionality of the deformation and strain meshes using, e.g., principal component analysis (PCA). Then, perform \emph{surrogate learning} by employing supervised learning and the reduced dataset (from the previous step) to obtain a DTP that predicts the reduced plastic strain from the reduced deformation.
\end{itemize}

A variety of tools and methods are available to support DT engineering across its four phases. For phase~\ref{dt-phi-1}, various tools exist, including Abaqus\footurl{https://www.3ds.com/products/simulia/abaqus}, Ansys Maxwell\footurl{https://www.ansys.com/products/electronics/ansys-maxwell}, OpenFOAM\footurl{https://www.openfoam.com/}.
Likewise, for phase~\ref{dt-phi-4}, numerous tools exist, such as Eclipse Ditto\footurl{https://eclipse.dev/ditto}, Azure Digital Twins\footurl{https://learn.microsoft.com/en-sg/azure/digital-twins/overview}, and iTwin.js\footurl{https://www.itwinjs.org/}. However, phases~\ref{dt-phi-2} and ~\ref{dt-phi-3} lack frameworks accounting for the particularities of DT engineering~\cite{gilSurveyOpensourceDigital2024}. 

\subsection{Contributions}
In this paper, we consider phases \ref{dt-phi-2} and~\ref{dt-phi-3}, as they both use data (either simulation or historical) to create surrogate models. 
In previous works, we have proposed the Function+Data Flow (FDF)~\cite{decontoFunction+DataFlowFramework2024}, a visual domain-specific language (DSL) to enable the construction of accurate and efficient DTs across diverse application domains using model-order reduction~\ref{dt-phi-2}, data assimilation~\ref{dt-phi-3}, or their combination. FDF treats functions as \emph{first-class citizens}, enabling their explicit and effective \emph{manipulation and reuse}. This is achieved by using a higher-order dataflow~\cite{priceHigherorderDataFlow2013,sivarajahTierkreisDataflowFramework2022,serotHoCLHighLevel2021}. Unlike a standard data flow, where nodes can only process data, a higher-order data flow allows nodes to accept functions as input and generate functions as output. We also recently introduced its concrete implementation in a tool called {\Builder}~\cite{decontoDesCartesBuilderTool2025}. 
By adopting \emph{model-driven engineering (MDE)} principles~\cite{michaelModelDrivenEngineeringDigital2025,combemaleHitchhikersGuideModelDriven2021}, which can facilitate DT engineering across phases~\ref{dt-phi-2} and~\ref{dt-phi-3}, FDF and {\Builder} aim to render DT engineering more accessible to domain experts. 

To assess the degree to which FDF and its implementation in {\Builder} facilitate the design of real-time DTPs and DTIs, we conducted an \emph{empirical user study}, which we report in this paper. The study was conducted as a controlled experiment (workshop) performed in an academic environment and using PhD students and researchers as subjects, and it aims to quantify the usability and adequacy of FDF and \Builder features. Subjects were asked to implement a solution to the material strain real-time DTP discussed in Fig~\ref{fig:pipe-strain-overview}. 
The results of the user study indicate that {\Builder} achieves \emph{good levels of usability}, as quantified using the standard System Usability Scale (SUS) questionnaire~\cite{brookeSUSQuickDirty1996}. This is remarkable since the tool is still in an early stage of development. Further data from our study allows us to identify strengths of our current approach as well as key areas for improvement going forward.

The second key contribution of this paper is to propose \emph{H-FDF (Hierarchical Function+Data Flow)}, an extension of FDF supporting iterative pipelines:
While FDF only supports strictly acyclic pipelines and is, thus, unable to support the specification of iterative processes, H-FDF addresses this limitation by explicitly defining modules that can be iterated internally. The iteration repeats until an explicitly defined stopping criterion is met, while the pipeline remains globally acyclic to facilitate its analysis and execution. We illustrate H-FDF on a dual training use case to efficiently learn dynamical systems with a residual learning approach~\cite{wuNonintrusiveModelCombination2024}.

\subsection{Paper organization}
The organization of this paper is as follows: 
Section~\ref{sec:related} introduces related research, placing FDF and {\Builder} in its broader research context. 
Section~\ref{sec:fdf} provides an overview of our Function+Data Flow DSL.
Section~\ref{sec:descartes-builder} recaps FDF's implementation in the {\Builder} tool, illustrating its practical usage in the material strain DTP (Fig.~\ref{fig:pipe-strain-overview}). 
Next, in Section~\ref{sec:empirical-study} we present our empirical user evaluation, including the methodology, results, and a discussion on their interpretation. 
Section~\ref{sec:hfdf} then presents key limitations of FDF and proposes the H-FDF extension to address them. 
Finally, Section~\ref{sec:conclusion} concludes the paper and identifies future research directions. 
 
\section{Related Work} \label{sec:related}

The integration of ML into DT workflows leverages several techniques, as we will now discuss.

\subsection{Multi-domain simulation} 

Tools such as Simulink\footurl{https://www.mathworks.com/products/simulink.html} offer the flexibility required for \emph{multi-domain physics simulation}~\ref{dt-phi-1}, which is essential in several DT applications~\cite{venkatesanHealthMonitoringPrognosis2019,Liu2023ADT,mehrabiAIDrivenDigitalTwin2024}.
For instance, constructing a DT of an electric motor involves modeling both mechanical (e.g, the rotor) and electrical (e.g., magnets) components: Simulink allows these components to be modeled using domain-specific paradigms and co-simulated within a unified data flow framework.
Nonetheless, Simulink lacks native support for ML model training and has limited support for interface typing and model reuse~\cite{benderSignatureRequiredMaking2015,jaskolkaSupportingModularitySimulink2020,mathworksItPossibleUse2009}. This makes it ill-suited for the integration of physics models with ML (hybrid modeling), an emerging trend in DT engineering~\cite{hartmannNextEvolutionMDE2019,combemaleEngineeringDigitalTwins}.
In contrast, FDF~\cite{decontoFunction+DataFlowFramework2024} and its implementation in {\Builder}~\cite{decontoDesCartesBuilderTool2025} address these limitations by providing a \emph{fully customizable} model reduction and instance-specialization pipeline which explicitly incorporates data-modeling aspects. 

\subsection{ML orchestration and DSLs} 

Scikit-learn~\cite{pedregosaScikitlearnMachineLearning2011} and PyTorch~\cite{paszkePyTorchImperativeStyle2019} are fundamental libraries for executing ML tasks and synthesizing ML models, as often required for fast DT learning~\ref{dt-phi-2} and DT instantiation~\ref{dt-phi-3}. \emph{ML workflow (MLOps)} tools such as Kedro~\cite{alamKedro2024}, MLFlow~\cite{chenDevelopmentsMLflowSystem2020}, and Metaflow\footurl{https://metaflow.org/} leverage these libraries to systematically represent complex pipelines, facilitating experiment tracking, model versioning, and performance monitoring. 
However, these tools generally focus on generating a single ML model, which remains implicit in the pipeline~\cite{lwakatareDataScienceDriven2020}. Thus,
many actions and the resulting ML model itself remain implicit in the tool. Also, the interface is primarily textual with limited built-in functions.
Our approach addresses this by providing a dedicated \emph{function flow} to explicitly represent the combination, manipulation, and reuse of the various models, as necessary in digital twinning.

Another line of research involves defining new DSLs or enhancing existing ones to formalize the ML pipeline and potentially enable code generation~\cite{moinModeldrivenApproachMachine2022,radlerIntegrationMachineLearning2022,giner-miguelezDescribeMLToolDescribing2022,hartmannNextEvolutionMDE2019}. This includes approaches leveraging, e.g., the SysML system engineering framework~\cite{wilkingIntegratingMachineLearning2022,radlerIntegrationMachineLearning2022}, which can also help to decompose the ML pipeline in hierarchical components. While these works move toward a higher degree of abstraction, they typically maintain the ML models implicitly and lack robust validation features.

FDF, in contrast, introduces \emph{implicit typing}, which can, transparently to the user, identify misconnections in the pipeline at specification time, ensuring the pipeline is structurally correct. Furthermore, the {\Builder} interface allows users to fully design and parametrize the pipeline, execute it, and validate the resulting models. At last, H-FDF enables the hierarchical decomposition of ML pipelines with the further capability of iterating until a suitable model is attained.

\subsection{Visual workflow environments}
Visual workflow environments such as Orange~\cite{demvsar2013orange}, RapidMiner\footurl{https://altair.com/altair-rapidminer} provide accessible graphical workflows that can enable domain experts to solve domain-specific problems more effectively~\cite{oakesBuildingDomainSpecificMachine2024}. Among these, we find \emph{KNIME (Konstanz Information Miner)}~\cite{bertholdKNIMEKonstanzInformation2009}, arguably the closest tool to our work due to its DAG-based pipeline representation and its flexible workflow. However, FDF and its hierarchical extension (H-FDF) differ from KNIME in two main aspects of their design philosophy:
\begin{itemize}
    \item \emph{Typing:} While FDF adopts an \emph{implicit typing} approach to represent the pipeline data and functions uniformly, KNIME adopts an explicit and nominal type system: a detailed comparison is provided in Sect.~\ref{sec:KNIME}. 
    \item \emph{Loops:} KNIME loops \emph{implicitly couple} the data flow and the control flow, resulting in a large (and arguably confusing) catalog of nodes. In contrast, H-FDF explicitly distinguishes these flows using hierarchical connections, thus clarifying the execution semantics and simplifying the evolution of the loop iteration strategy (see Sect.~\ref{sec:loops-knime-vs-fdf}).
\end{itemize}
 
\section{FDF Overview}
\label{sec:fdf}

This section presents the Function+Data Flow (FDF) domain-specific language~\cite{decontoFunction+DataFlowFramework2024}. FDF specifically targets the demands of the fast DT learning~\ref{dt-phi-2} and DT instantiation~\ref{dt-phi-3} phases, which require combining multiple co-dependent ML models. 
To address this, FDF represents functions (ML models) as \emph{first-class citizens}, allowing them to be passed as inputs, produced as outputs, and reused as required. FDF also defines three primitive boxes (\Processor, \Coder, and \Trainer) that map directly to the key ML tasks in~\ref{dt-phi-2} and~\ref{dt-phi-3}: model reuse, reduced-order modeling, and surrogate modeling.

We now discuss the key FDF concepts. For further details, we refer the reader to~\cite{decontoFunction+DataFlowFramework2024}. 

\begin{figure*}[t!]
    \centering
    \includegraphics[width=\textwidth]{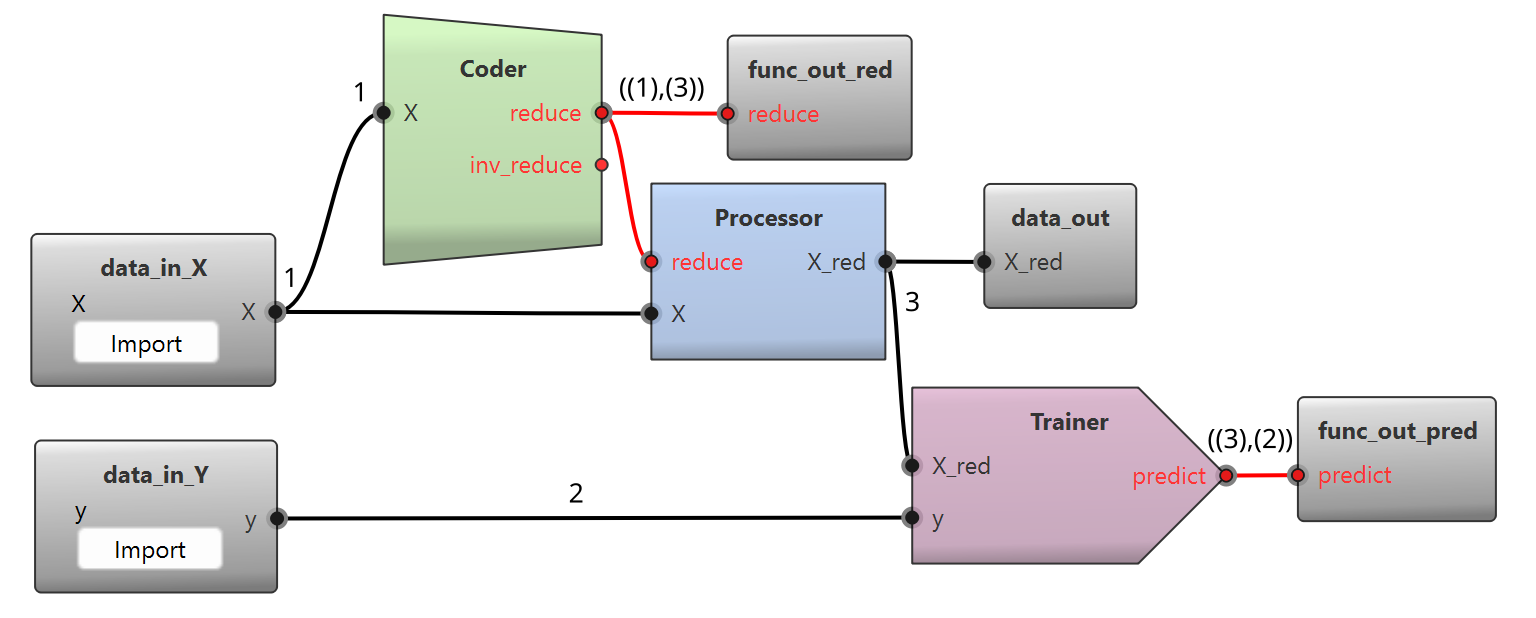}
    \caption{Minimal FDF pipeline implemented in {\Builder}. The number next to the ports indicates a possible inferred implicit type}
    \label{fig:fdf-minimal}
\end{figure*}

\subsection{FDF pipeline, graph and ports} \label{sec:fdf-pipe-ports}

An \emph{FDF pipeline} $\FDpipe$ is defined as a directed graph $G(\FDpipe) = (\FDports, \FDedges)$, where $\FDedges$ models the flow between the ports in $\FDports$.
In contrast to a typical data flow language, which only has a single type of port (representing data), FDF defines two explicit types of ports:
\begin{itemize}
    \item \emph{function ports} (depicted in \emph{\color{red}red}), which send/receive a \emph{learned function}, and
    \item \emph{data ports} (depicted in \emph{black}), which send/receive data batches.
\end{itemize}

Each FDF port $\FDport$ is either an \emph{{\IP}} or \emph{{\OP}}, and belongs to a single \emph{FDF box} $\FDbox=\portbox(\FDport)$. $\FDedges$ is {\bf acyclic}.  
For $p'$ an \IP, there is only one $p$ such that 
$(p,p') \in \FDedges$. Further $p$ is an \OP, of the same type as $p'$ 
(function or data). 
Lastly, if $p'$ is an {\OP}, then $(p,p') \in \FDedges$ iff $\portbox(p)=\portbox(p')$
and $\FDport$ is an {\IP}.

\subsection{FDF boxes} \label{sec:fdf-boxes}

The boxes correspond to the tasks to be executed. There are three box types: ({\Processor}, {\Coder}, and {\Trainer}) that can be used to specify the ML pipelines for digital twinning:
\begin{itemize}
    \item {\Processor} is used for data processing tasks, especially calling the functions learned by the other boxes,
    \item {\Coder} is for unsupervised learning using algorithms such as principal component analysis (PCA), and
    \item {\Trainer} is for supervised learning using, e.g., a neural network architecture learned with stochastic gradient descent (SGD). 
\end{itemize}

The data and function dependencies are given by gray boxes, namely \DataIO and \FuncIO. Note that each FDF box is visually represented by a unique geometric shape and color: the \Processor adopts a rectangle commonly used in workflow diagrams, the \Coder uses a trapezoid commonly associated with dimensionality reduction (e.g., auto-encoders), and the \Trainer narrows down to a single point since it outputs only one function. Distinct colors are also assigned to each FDF box, providing redundant visual encoding that can enhance their discriminability~\cite{moodyPhysicsNotationsScientific2009}.

\subsection{Minimal example}
We now describe a minimal FDF pipeline implementing a simple real-time DTP~\ref{dt-phi-2}, as implemented in {\Builder}, in Fig.~\ref{fig:fdf-minimal}\footnote{Available as \emph{minimal.dcb} in the paper artifact.}. This pipeline proceeds as follows: 
\begin{enumerate}
    \item \emph{Data loading.} Use two \DataIO gray boxes (\verb|data_in_X| and \verb|data_in_Y|) to load \verb|X| and \verb|y|.
    \item \emph{Unsupervised ML.} Employ a \Coder box to compute a reduced space for the input data \verb|X| and return its associated \verb|reduce| function.
    \item \emph{Data encoding.} A \Processor box encodes \verb|X| as \verb|X_red| using the \verb|reduce| function.
    \item \emph{Supervised ML.} Last, a \Trainer box learns a \verb|predict| function that takes \verb|X_red| as input and outputs an estimation of \verb|y|.
\end{enumerate}
The pipeline also exports the functions \verb|reduce| and \verb|predict| that can be used to exploit the resulting real-time DTP with new data and to perform PT interaction~\ref{dt-phi-4}.

\subsection{Implicit typing} \label{sec:implicit-typing}

In FDF, the concept of \emph{implicit typing} is a core mechanism that ensures structural consistency in digital twinning pipelines. By treating functions as first-class citizens, FDF enables \emph{automatic inference} of implicit types for both data and functions within the pipeline. 
We now summarize the rationale for implicit typing, and we describe how to infer these types based on the FDF syntax and semantics by forward propagation in the FDF graph. For further details, we refer to the original FDF paper~\cite{decontoFunction+DataFlowFramework2024}. 

\subsubsection{Rationale}

Implicit typing endows the FDF pipeline with advantages commonly associated with statically typed languages, e.g., being less prone to errors and easier to maintain~\cite{bognerTypeNotType2022,rayLargeScaleStudy2014}. Because the typing is \emph{implicit}, users do not have to manually define explicit types, which can be laborious~\cite{oreAssessingTypeAnnotation2018}. More importantly, explicit typing is often infeasible at design time in the case of DT design, where pipeline functions are learned dynamically.
For instance, when applying PCA to obtain a reduced basis that preserves 99\% of the original dataset variance, the explicit output type (i.e., number of dimensions) depends on the training data: explicit typing {\em at design time} (before the training data has been processed) would not be feasible in this case. 

Further, FDF implicit types represent the pipeline \emph{semantics} (e.g., "the reduced space of dataset $X$"), akin to \emph{structural typing}. In this paradigm, a type $S$ is a subtype of $T$ if $S$ provides the structure (i.e., interface) required by $T$ (e.g., required members and functions)~\cite{cardelliStructuralSubtypingNotion1988}. In contrast, \emph{nominal typing} treats $S$ as a subtype of $T$ only if this relationship is explicitly declared~\cite{pierce2002types}. 

Note that these two paradigms sit on opposite ends of a typing design spectrum. While languages like Java or C++ are almost purely nominal, others like TypeScript are primarily structural~\cite{bierman2014understanding}. Meanwhile, languages like Scala and Maude blend these two approaches: Scala, for example, has a nominal core but provides native, optional support for structural types\footurl{https://docs.scala-lang.org/scala3/reference/changed-features/structural-types.html}. In this landscape, FDF is strongly structural, and it eliminates the need for signature or interface declarations. Instead, FDF implicit types are derived solely based on the FDF graph (Sect.~\ref{sec:fdf-pipe-ports}) and the DSL boxes.

\subsubsection{Type definition}

We distinguish two categories of implicit types: data types and function types. 

A \emph{data type} is associated with each {\DP} and represented by a unique identifier, the source {\DP} ID.
Formally, for each \DP\ $p$, there exists a 
\DP\ $p'$ with $\typeop(p)=p'$. A priori, two unrelated {\DP}s have different types, though users can specify type equality.

A \emph{function type} is similarly associated with each {\FP} and defined as a tuple matching the sequence of input data types to the sequence of output data types. Thereby, a function $f$ with $k$ inputs and $k'$ outputs is assigned the implicit type
$\typeop(f) = ((\type_{1}, \ldots, \type_{k}),(\type_{k+1},\ldots,\type_{k+k'})),$
where $\type_{i \leq k}$ is the implicit type of the $i$-th input, and $\type_{k+i}$ is the implicit type of the $i$-th output of $f$ for $i\leq k'$.

\subsubsection{Type inference, propagation and checking}
\label{sec:fdf-type-infer}

The data types and function types are automatically inferred in FDF based on the box and then \emph{forward propagated} between ports. This process occurs in three distinct phases, which we summarize in the following (refer to~\cite{decontoFunction+DataFlowFramework2024} for details). 

\paragraph{Phase I: initialization}
Types are established at the pipeline boundary (\DataIO and \FuncIO boxes). Each {\OP} of these boxes is assigned a unique default type $\typeop(i) \gets i$. For instance, in Fig.~\ref{fig:fdf-minimal}, \verb|X| {\ODP} belonging to \verb|data_in_X| \DataIO box could be given type `1'.

\paragraph{Phase II: propagation}
Types flow along the edges of the graph with every input port $\FDport_{in}$ copying the implicit data type from its corresponding {\OP} $\FDport_{out}$: $\typeop(\FDport_{in})\gets \typeop(\FDport_{out})$. In Fig.~\ref{fig:fdf-minimal}, this means that the {\IDP} \verb|X| of Coder and Processor would also have type `1'.

\begin{figure}[t!]
    \centering
    \includegraphics[width=0.95\columnwidth]{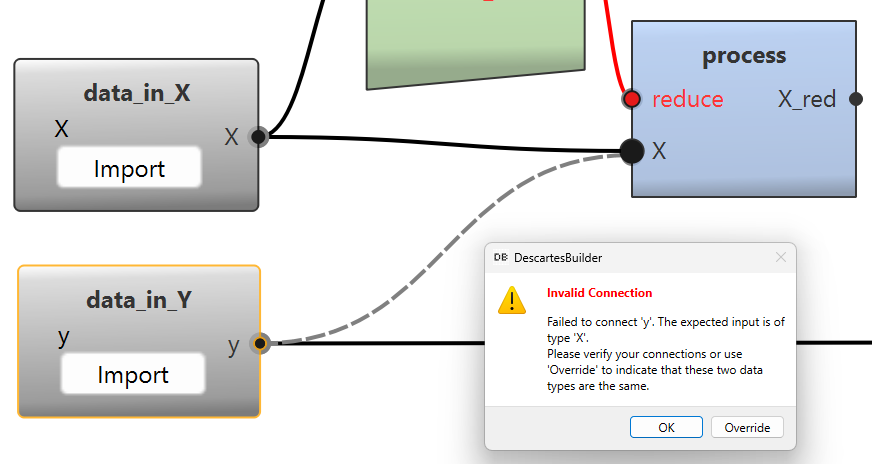}
    \caption{Example of FDF's implicit type warning as implemented in {\Builder} when attemping to connect {\tt y} to the function {\tt reduce} that expects an inputs with same implicit type as {\tt X} in the FDF pipeline of Fig.~\ref{fig:fdf-minimal}}
    \label{fig:fdf-type-warning}
\end{figure}

\paragraph{Phase III: implicit type inference}

The final step computes types for a box's output ports based on the box class (\Coder, \Trainer, or \Processor):

\begin{itemize}
    \item For \Coder, this consists of creating new "fresh" types for the compressed representations. For instance, in Fig.~\ref{fig:fdf-minimal}, the type of \verb|reduce| could be `$((1),(3))$'. 
    \item Given a function $f$ whose type is $\typeop(f) = ((\type_{1}, \ldots, \type_{k}),(\type_{k+1},\ldots,\type_{k+k'}))$, a \Processor will propagate the data types $\type_{k+1},\ldots,\type_{k+k'}$ unless a warning is triggered~\cite{decontoFunction+DataFlowFramework2024}. In Fig.~\ref{fig:fdf-minimal}, we thus propagate type `3'. A warning would be triggered if, e.g., we attempted to connect \verb|y| instead of \verb|X| to the \Processor box (Fig.~\ref{fig:fdf-type-warning}). 
    \item A \Trainer assigns a function type based on the split of $(X,Y)$ training data. Hence, in Fig.~\ref{fig:fdf-minimal}, the function \verb|predict| could have type $((3),(2))$.
\end{itemize}

\subsection{Comparison with KNIME} 
\label{sec:KNIME}

\begin{figure*}[t]
    \centering

    \begin{subfigure}[b]{0.95\textwidth}
        \centering
        \includegraphics[width=\textwidth]{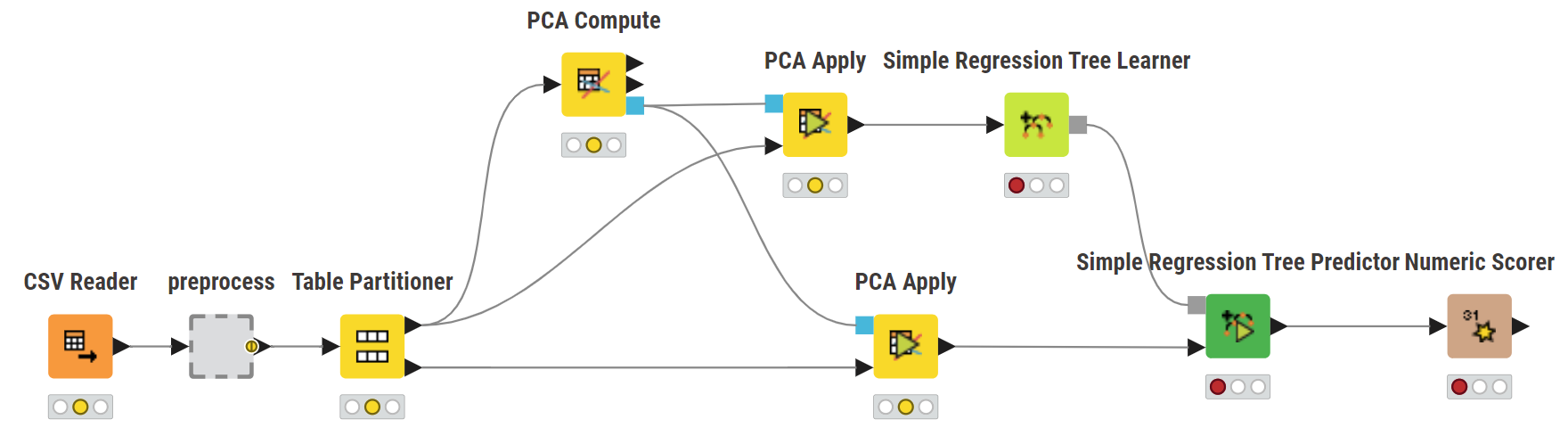}
        \caption{A KNIME pipeline to learn a decision tree. Observe that, before the pipeline execution, some nodes are marked in {\color{red}{red}} to indicate that they are not yet fully configured. This is because the number of dimensions generated by the "PCA Apply" node is unknown in advance. Hence, the choice of target used by the Learner needs to be postponed until the prior nodes are executed}
        \label{fig:1-housing-knime}
    \end{subfigure}
    
    \vspace{1em}

    \begin{subfigure}[b]{0.95\textwidth}
        \centering
        \includegraphics[width=\textwidth]{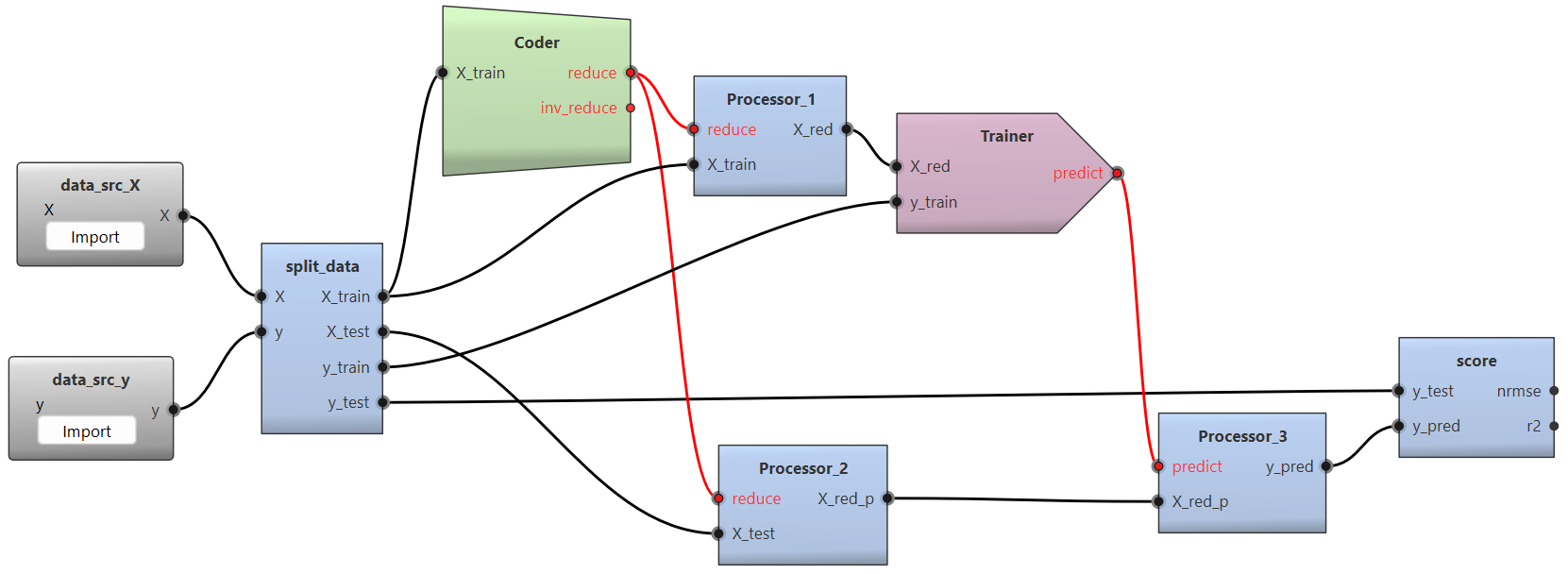}
        \caption{Equivalent FDF pipeline implemented within {\Builder}. All boxes are implicitly typed, and all inputs and outputs are explicitly observed in the FDF graph}
        \label{fig:1-housing-builder}
    \end{subfigure}

    \caption{Comparison of FDF pipeline in {\Builder} (top) and equivalent KNIME pipeline (bottom)}
    \label{fig:housing-comp}
\end{figure*}

KNIME (Konstanz Information Miner)~\cite{bertholdKNIMEKonstanzInformation2009} is a visual workflow for data mining tasks. KNIME is an established tool that has been developed for more than 20 years, with an open-source architecture and good support for adding new nodes via a plug-in system. Similar to FDF, the KNIME pipeline is defined as a directed acyclic graph with a series of interconnected nodes (analogous to FDF's boxes). However, KNIME and FDF have key differences in their design philosophy, as we now discuss. 

The first key distinction is in the \emph{pipeline granularity}. KNIME provides an extensive repository of generic nodes for data-mining and ML tasks. While flexible, this can increase the user's cognitive load as the user needs to navigate and choose the adequate node for the task at hand. Conversely, FDF introduces a minimal set of domain-specific boxes (\Processor, \Coder, \Trainer) tailored for DT engineering tasks. This abstraction allows the pipeline to be described with high-level concepts aligned with user intent rather than low-level details of specific algorithms.

Another key difference concerns \emph{typing}. KNIME relies on an \emph{explicit and nominal} type system for its ports.
Specifically, KNIME types need to support explicit assignment, as confirmed by inspecting its source code\footurl{https://github.com/knime/knime-core/blob/7991b4a31ebeaa67101d210eed091de9c423c09a/org.knime.core/src/eclipse/org/knime/core/node/workflow/WorkflowManager.java\#L1747}.
In contrast, FDF employs \emph{implicit typing} (Sect.~\ref{sec:implicit-typing}), which is analogous to the notion of \emph{structural typing}~\cite{cardelliStructuralSubtypingNotion1988}.
In FDF, this is implemented by using a single and general function type that applies uniformly to all functions and ML models.

Furthermore, ML models in KNIME are encapsulated with custom types such as "Regression Tree Model" for a regression tree or "Neural Network Model" for a neural network.Although some generalization is possible by, for example, representing the models with types derived from the Predictive Model Markup Language (PMML), a \emph{tight coupling} remains between a model’s internal representation and the nodes that process it. As a result, the platform typically requires a 1:1 correspondence between a specific Learner node and its associated Predictor node (e.g., "PCA Compute" and "PCA Apply" in Fig.~\ref{fig:1-housing-knime}). 

FDF breaks this coupling (Fig.~\ref{fig:1-housing-builder}) by representing these operations through generic components: "PCA Compute" is mapped to a generic \Coder box configured with the PCA parameters, while "PCA Apply" is handled by a generic \Processor box that takes as input the {\tt reduce} function computed by \Coder.

The main consequence of KNIME's explicit typing is that \emph{swapping ML models can be cumbersome}. 
As shown in the "California housing"~\cite{kelleypaceSparseSpatialAutoregressions1997} regression example (Fig.~\ref{fig:housing-comp}), replacing a "Regression Tree" with a "Neural Network" in KNIME (Fig.~\ref{fig:1-housing-knime}\footnote{Available as \emph{knime-vs-fdf/housing-knime.knwf} in the paper artifact.}) requires \emph{five distinct user actions}:

\begin{enumerate}
    \item Change the Learner from "Regression Tree" to "Neural Network",
    \item Reconnect the Learner to the training data,
    \item Redefine the correct "target column" in the Learner node,
    \item Replace the "Regression Tree Predictor" node with a "Neural Network Predictor", and
    \item Reconnect the predictor to the test data.
\end{enumerate}

In contrast, in FDF (Fig.~\ref{fig:1-housing-builder}\footnote{Available as \emph{knime-vs-fdf/housing-fdf.dcb}}), the algorithm is treated uniformly as a \Trainer box parameter, allowing models to be swapped with a \emph{single click} in the {\Builder} implementation.

\section{The DesCartes Builder Tool}
\label{sec:descartes-builder}

To demonstrate the practical utility of the FDF framework, we developed {\Builder}, an open-source tool designed to systematize the engineering of ML-based DT pipelines~\cite{decontoDesCartesBuilderTool2025}. The tool provides a visual modeling environment where domain experts (e.g., an engineer responsible for designing a DT for a manufacturing industry) can specify, execute, and validate DT synthesis workflows without deep programming expertise by leveraging FDF (Sect.~\ref{sec:fdf}). Key features include: \begin{itemize}
    \item Graphical modeling of the DT synthesis pipeline using the FDF boxes, 
    \item Explicit modeling of both data flows and function flows and their interconnections,
    \item Automatic code generation from FDF specifications, 
    \item Execution of the DT synthesis pipeline, producing data artifacts and learned models, 
    \item Static validation of the DT synthesis pipeline using \emph{implicit typing}, enabled by FDF's treatment of functions as first-class citizens, and
    \item Built-in support for validating synthesized ML models and data artifacts.
\end{itemize}

{\Builder} is released as an \emph{open-source} software, enabling extensibility and community-driven development~\cite{decontoDesCartesBuilderV022}. The following sections detail the system architecture (Sect.~\ref{sec:builder-archi}) and illustrate the implementation of a real-time DTP~\ref{dt-phi-2} within the tool (Sect.~\ref{sec:dtp-material-strain}).

\subsection{Tool's Architecture} \label{sec:builder-archi}

\begin{figure}[b!]
    \centering
    \includegraphics[width=\linewidth]{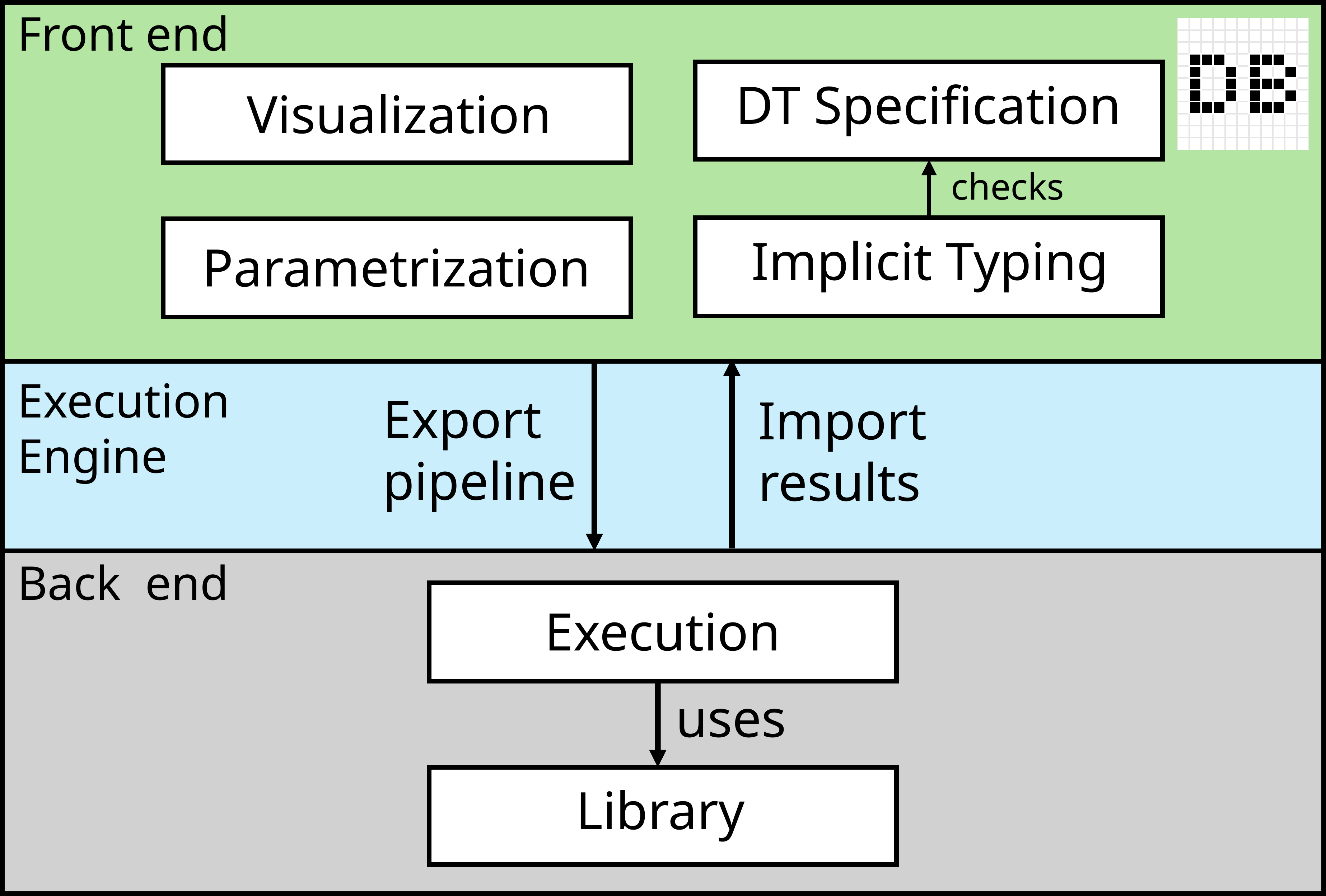}
    \caption{Depiction of the {\Builder} architecture}\label{fig:builder_architecture}
\end{figure}

\begin{figure*}[t!]
    \centering
    \includegraphics[width=\textwidth]{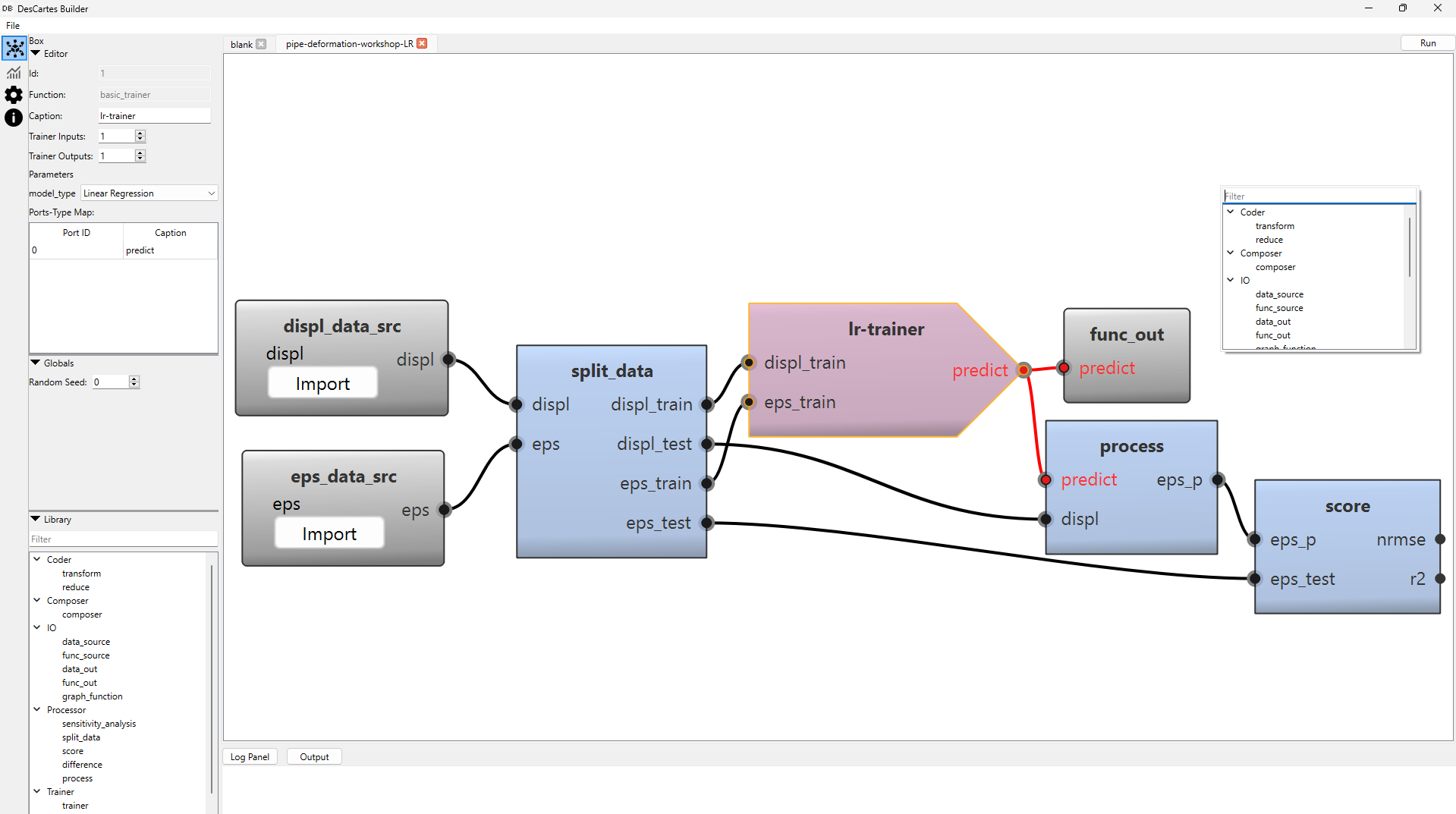}
    \caption{{\Builder} user interface showing the canvas~\ref{ui-canvas}, the library selector~\ref{ui-lib}, and the parameter editor~\ref{ui-param}. A simplified implementation of the material strain prediction real-time DTP (Fig.~\ref{fig:pipe-strain-overview}) is displayed. It uses a \texttt{split\_data} block to partition displacement ({\tt displ}) and plastic strain ({\tt eps}) into training ({\tt \_train}) and test ({\tt \_test}) sets. The \Trainer then learns a function {\color{red}\texttt{predict}} from ({\tt displ\_train}, {\tt eps\_train}), which a \Processor then applies to {\tt displ\_test} to estimate {\tt eps\_p}. 
    Last, a {\tt score} box compares {\tt eps\_p} against the test data {\tt eps\_test}, outputting metrics such as NMRSE and $R^2$. The function {\color{red}\texttt{predict}} is exported to constitute a real-time DTP.}
    \label{fig:builder-startup}
\end{figure*}

{\Builder} is structured into two primary components (Fig.~\ref{fig:builder_architecture}): a graphical front end and an execution back end. This ensures a clear separation between the user interface (UI) and the core computational logic. The interaction between these components is mediated by an \emph{abstract execution engine}, providing a decoupled communication layer.

\subsubsection{Front end: user interface}

{\Builder} front end offers a unified interface for the specification, execution, and validation of ML-based DT pipelines. Developed in C++ using the Qt framework and the QtNodes library~\cite{pinaevQtNodesNodeEditor2024}, the UI (Fig.~\ref{fig:builder-startup}) comprises the following core components:
\begin{enumerate}[label={($C_\arabic*$)}]
    \item\label{ui-canvas} a central \emph{canvas} for modeling DT pipelines using FDF, allowing users to create boxes and to interconnect their ports through drag-and-drop. Function and data connections are depicted in \textbf{\color{red}red} and \textbf{black}, respectively, for clear differentiation. 
    \item\label{ui-lib} a \emph{library selector} (canvas pop-up and left pane) for inserting FDF boxes,
    \item\label{ui-param} a \emph{parameter editor} (left pane) for modifying the parameters of FDF boxes (e.g., adjusting the data encoding learned by the \Coder).
    \item A \texttt{Run} button (top-right) for triggering the \emph{pipeline synthesis}.
    \item A \verb|Log Panel| and \verb|Output| panel (bottom) for displaying execution traces.
    \item\label{ui-chart} A \emph{chart viewer} (\includegraphics[height=0.8\baselineskip]{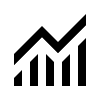}) for visualizing the results of pipeline execution using charts and validation metrics visible by selecting appropriate boxes (e.g., \verb|score| Processor).
\end{enumerate}

\begin{figure*}[t!]
    \centering
    \includegraphics[width=\linewidth]{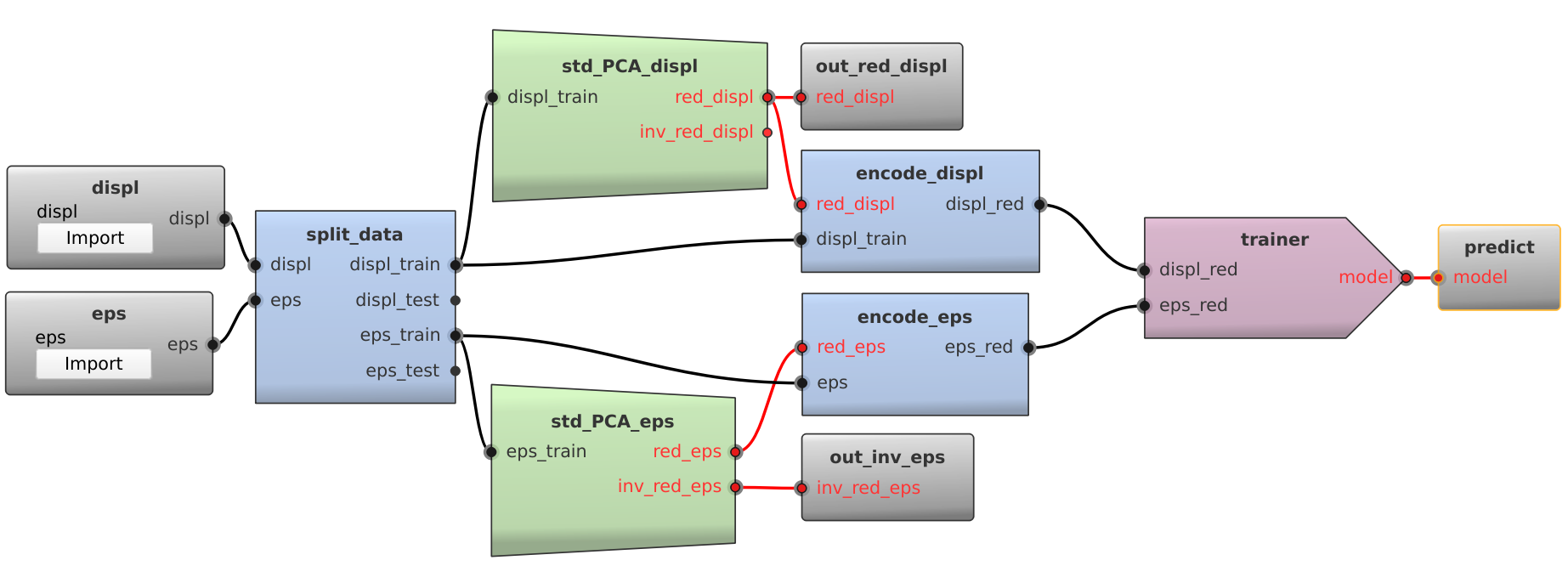}
    \caption{Implementation of the material strain reduced-order model (ROM). It uses two \Coder boxes to learn the dimensionality reduction basis, two Processors to obtain the reduced displacement (\texttt{displ\_red}) and reduced strain (\texttt{eps\_red}), and a Trainer to learn a surrogate model from \texttt{displ\_red} to \texttt{eps\_red}. Evaluation is performed using a \texttt{score} and a \texttt{sensitivity\_analysis} Processor box (adapted from~\cite{decontoDesCartesBuilderTool2025})}\label{fig:pipe-deformation-train}
\end{figure*}

The \emph{pipeline synthesis} begins by mapping the graphical FDF model into a back-end configuration. The execution engine then orchestrates the execution of the back-end, training, and validating ML models as needed. Upon completion, it collects the results, presenting them both in the \verb|Output| panel and in the \emph{chart viewer}.

\subsubsection{Back end: execution engine}

The back end, implemented in Python as a Kedro~\cite{alamKedro2024} plugin, has two main responsibilities:
\begin{enumerate}
    \item \emph{Pipeline orchestration}: Loads the input data, determines the execution order and dependencies within the FDF graph (Sect.~\ref{sec:fdf-pipe-ports}) and saves the outputs as required.
    \item \emph{Library implementation}: Provides the concrete logic for FDF boxes, which was implemented leveraging established ML libraries such as Scikit-learn~\cite{pedregosaScikitlearnMachineLearning2011} and PyTorch~\cite{paszkePyTorchImperativeStyle2019}.
\end{enumerate}

\subsection{Real-time DT prototype for material strain prediction}
\label{sec:dtp-material-strain}

We now showcase the application of {\Builder} to model and validate a real-time DTP for material strain prediction (Fig.~\ref{fig:pipe-strain-overview}). Recall that the goal is to predict a structure's plastic strain from an observed deformation~\cite{chabodDigitalTwinFatigue2022}. 

\subsubsection{Modeling} 

As discussed in Sect.~\ref{sec:introduction}, to design this DTP, we first need a high-fidelity DT~\ref{dt-phi-1} of the system under consideration, which we obtained with the assistance of a collaborator. However, since each simulation requires around one hour of runtime, they are impractical for a real-time DTP. Consequently, we adopted a design of experiments (DoE) strategy to systematically generate a representative dataset of the material's behavior. 

Using the DoE data, we developed a reduced-order model (ROM) of the structure in {\Builder} that correlates the deformation (\verb|displ|) and the material plastic strain (\verb|eps|). The synthesis pipeline, illustrated in Fig.~\ref{fig:pipe-deformation-train}\footnote{Available as \emph{pipe-deformation-train.dcb} in the paper artifact.}, takes two data inputs and produces three function outputs. 
The inputs are: (1) \verb|displ|, the structure's observed \emph{displacement}, and (2) \verb|eps|, the associated plastic strain. Both \verb|displ| and \verb|eps| have 276 samples and 2835 features. 
The outputs are: (1) \verb|model|, the resulting surrogate model, (2) \verb|red_displ| to encode (reduce) the displacement into a reduced basis (required as input to the surrogate model), and (3) \verb|inv_red_eps| to decode the output from the surrogate into the full-dimensional physical space.

\begin{figure*}[t!]
    \centering
    \includegraphics[width=\linewidth]{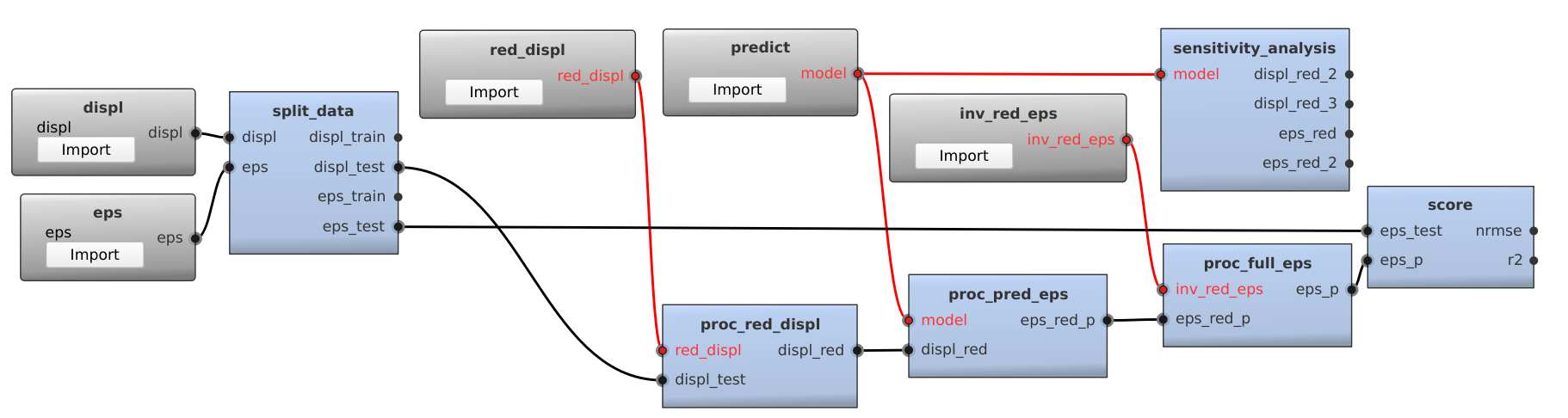}
    \caption{Testing pipeline for the material strain ROM}
    \label{fig:pipe_deformation_test}
\end{figure*}

We then implement the DTP in {\Builder} as follows:
\begin{enumerate}
    \item \emph{Data loading and train/test split:} The input \verb|displ| and \verb|eps| are loaded using a \verb|DataIO| box, and a \verb|split_data| \Processor splits \verb|displ| and \verb|eps| between train and test sets.
    \item \emph{Learn reduced basis:} Two \Coder boxes (\verb|std_PCA_displ| and \verb|std_PCA_eps|) are used to learn from the training data the encoding (\verb|red_displ| and \verb|red_eps|) and decoding (\verb|inv_red_displ| and \verb|inv_red_eps|) to/from the reduced basis using PCA for each input. The \Coder box is parametrized to use standardization followed by PCA such that 99.9\% of the original variance is maintained in the reduced basis.
    \item \emph{Data reduction:} To obtain the reduced data, two \Processor boxes (\verb|encode_displ| and \verb|encode_eps|) encode the full-dimensional data with the learned encodings (\verb|red_displ| and \verb|red_eps|), resulting in \verb|displ_red| and \verb|eps_red|, the reduced displacement and plastic strain. Hence, we reduce the dimensionality of the deformation and strain meshes from $2835$ dimensions to $\approx10$ dimensions.
    \item \emph{Surrogate learning:} A \Trainer learns the surrogate \verb|model| to estimate \verb|eps_red| from \verb|displ_red|. It is parametrized to learn a neural network of 2 layers with 50 nodes each, using the Adam optimizer, with a learning rate of $0.001$, and for 1000 iterations. The resulting \verb|model| is outputted.
\end{enumerate}

Note that {\Builder} facilitates the modeling of this pipeline in three ways. First, it allows describing visually the data flow necessary to learn a function. Second, it enables the easy export of functions using the gray output boxes (\verb|out_red_displ|, \verb|out_inv_eps|, \verb|predict|). Third, it supports FDF's implicit typing, which can prevent user mistakes by identifying misconnections at specification time (Sect.~\ref{sec:implicit-typing}). For example, it can prevent users from feeding \verb|displ| instead of the expected \verb|eps| to the \verb|encode_eps| Processor. 

\subsubsection{Validation and exploitation}

\begin{figure*}[t]
    \centering
    \begin{subfigure}[b]{0.45\textwidth}
        \centering
        \includegraphics[height=0.25\textheight]{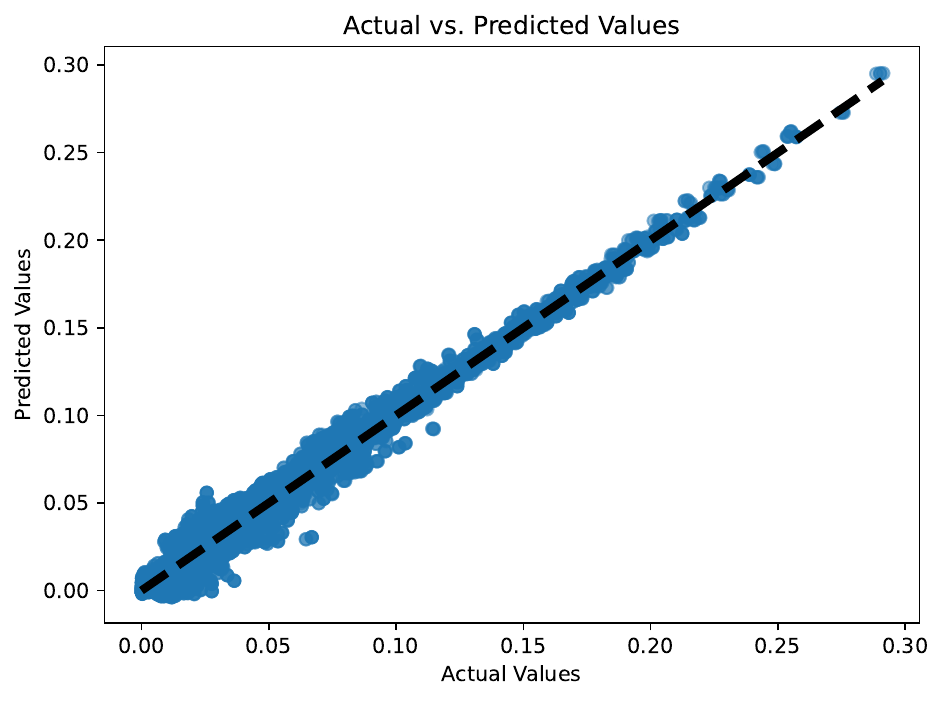}
        \caption{Actual vs.\ predicted values plot}\label{fig:pipe_deformation_actual_vs_pred}
    \end{subfigure}
    \hfill
    \begin{subfigure}[b]{0.45\textwidth}
        \centering
        \includegraphics[height=0.25\textheight]{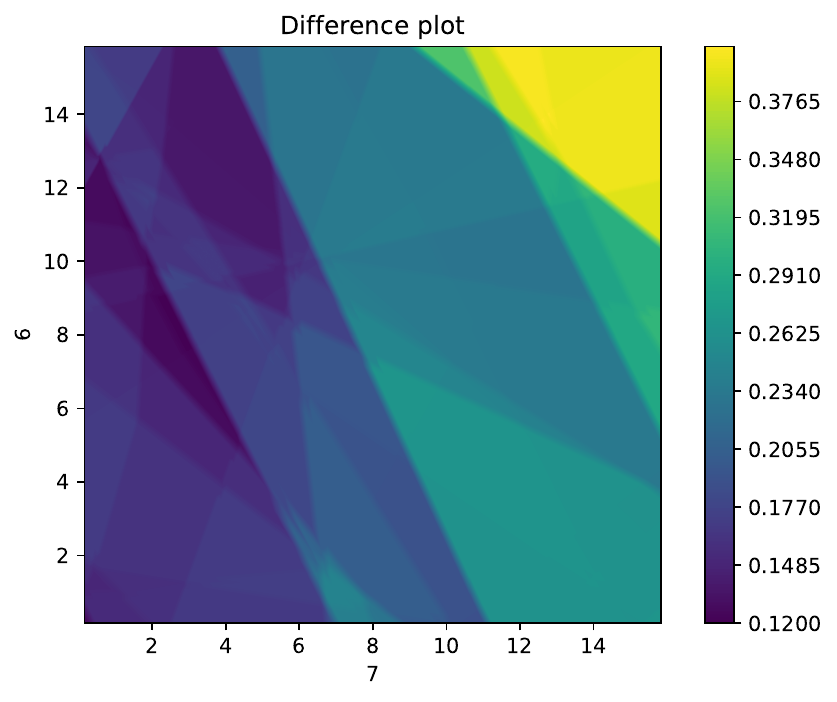}
        \caption{Sensitivity analysis plot}\label{fig:pipe-deformation-diff-plot}
    \end{subfigure}
    \caption{Validation results for the material strain prediction use case}\label{fig:pipe_deformation_valid}
\end{figure*}

Once the different functions are learned, the DTP can then determine the plastic strain from, e.g., actual 3D images of a deformation. We can describe the exploitation pipeline in FDF and implement it in {\Builder} using a series of \Processor boxes that leverage the learned functions. 
We illustrate the process by demonstrating the pipeline on the test set. The testing pipeline (Fig.~\ref{fig:pipe_deformation_test}\footnote{Available as \emph{pipe-deformation-test.dcb} in the paper artifact.}) proceeds as follows:

\begin{enumerate}
    \item \emph{Data loading:} Load the \verb|displ| and \verb|eps| data and obtain the test data using a \verb|split_data| box.
    
    \item \emph{Generate predictions:} Apply the encoder \verb|red_displ| to obtain the reduced displacement \verb|displ_red|. The resulting \verb|displ_red| can be used as input to the strain model \verb|predict|, resulting in a reduced \verb|eps_red_p|. Then, we decode \verb|eps_red_p| using \verb|inv_red_eps| into a full dimensionality strain map \verb|eps_p|.
    
    \item \emph{Scoring:} We perform the standard ML scoring using a \verb|score| \Processor which computes NMRSE and $R^2$ between \verb|eps_red| and \verb|eps_red_p|. It also generates, among others, a plot of actual vs. predicted values (Fig.~\ref{fig:pipe_deformation_actual_vs_pred}).

    \item \emph{Model analysis:} Since functions in FDF are first-class citizens, we can also specify an algorithm whose input is a function. For instance, {\Builder} supports a \verb|sensitivity_analysis| Processor (Appendix~\ref{sec:sensitivity-analysis}) that can be used to analyze the ML surrogate \verb|model|. This box finds two $\varepsilon$-close displacements (\verb|displ_red_s1|, \verb|displ_red_s2|) resulting in two plastic strains (\verb|eps_red_s1| and \verb|eps_red_s2|) which are significantly different. It also produces a diagram (Fig.~\ref{fig:pipe-deformation-diff-plot}) to visualize the gradient over sensitive dimensions, for validation purposes. In this plot, the X-Y axis (7/6) indicates the input features of \verb|disp| used for sensitivity analysis, and the heatmap indicates the difference for a chosen \verb|eps| output feature (0 in this case). 
\end{enumerate}

Executing this pipeline in {\Builder} yields an ML surrogate for \emph{model order reduction} with good performance ($R^2$ score of 0.8). As shown in Fig.~\ref{fig:pipe_deformation_actual_vs_pred}, predicted values align closely with the ground truth along the ideal $45^\circ$ dashed line. 
However, localized data scarcity persists, as indicated by the higher residuals in the upper-right corner of Fig.~\ref{fig:pipe-deformation-diff-plot}. This could be addressed, e.g., by improving the DoE strategy. 
Last, we can also export the results from \verb|score|, \verb|sensitivity_analysis|, and \verb|eps_pred| for expert review. 

Note that the complexity of this pipeline is justified by its accuracy. Indeed, a simplified alternative (Fig.~\ref{fig:builder-startup}) utilizing the same training set achieves an $R^2$ score of only 0.5, representing a 150\% increase in unexplained variance.

\section{Empirical Study on FDF and DesCartes Builder} \label{sec:empirical-study}

To evaluate the efficacy of the Function+Data Flow (FDF) framework and its associated modeling environment, {\Builder}, we conducted a controlled user study~\cite{deContoSoSymArtifact}. The primary \emph{objective} was to assess whether the proposed methodology facilitates the development of Digital Twins (DTs) for our target users, e.g., domain experts responsible for designing a DT for a manufacturing industry. Our evaluation is guided by two research questions (RQs):
\begin{enumerate}[label={($RQ_\arabic*$)}]
    \item\label{rq1-usab} \emph{Usability}: To what extent do users perceive {\Builder} as a usable for modeling a ML-based DT?
    \item\label{rq2-feat} \emph{Feature adequacy}:  Do users perceive {\Builder}'s current features as adequate for the DT modeling tasks encountered in this study? Which features are perceived as most and least effective by users? 
\end{enumerate}

\subsection{Study protocol}

The user study was conducted at NTU and CNRS@CREATE in Singapore, supervised by the first and second authors. Participants were recruited via internal mailing lists and provided informed consent (Appendix~\ref{app:informed-consent}) per ethical standards for empirical software engineering~\cite{singerEthicalIssuesEmpirical2002}. The research procedure was approved by the NTU Institutional Review Board (IRB-2025-1097).

The study was structured as two workshops in February 2026, each lasting approximately two hours and organized into three phases:
\begin{enumerate}
    \item \emph{Tutorial} ($\approx30$ min): Participants were introduced to {\Builder} via a guided walkthrough using the material strain prediction use case (Fig.~\ref{fig:pipe-strain-overview}). They implemented a baseline linear regression pipeline (Fig.~\ref{fig:builder-startup}\footnote{Available as \emph{user-study-material/pipe-deformation-workshop-LR.dcb} in the paper artifact.}) which achieves an intentionally low $R^2$ (0.5), leaving room for experimentation.
    \item \emph{Independent exploration} ($\approx60$ min): Participants were tasked with exploring the tool to improve the baseline pipeline independently. They were encouraged to 
    leverage the different features of the tool, e.g., changing the \Trainer algorithm or adding \Coder boxes for data compression.
    \item \emph{Evaluation} ($\approx30$ min): Participants completed a structured survey (Appendix~\ref{app:user-survey}) comprising:
    \begin{enumerate}
        \item demographic and background information; 
        \item a system usability scale (SUS) questionnaire~\cite{brookeSUSQuickDirty1996} to assess usability (RQ1);
        \item feature-specific usability ratings on a five-point Likert scale (1: Very demanding, 5: Very intuitive) to assess feature adequacy (RQ2).
        \item open-ended qualitative feedback regarding usability barriers and suggestions.
    \end{enumerate}
\end{enumerate}

\subsection{Evaluation instruments}\label{sec:usability-eval}

\begin{table}[t]
    \centering
    \caption{Questionnaire for {\Builder} usability (based on the SUS questionnaire~\cite{brookeSUSQuickDirty1996})}
    \label{tab:sus-builder}
    \begin{tabularx}{\columnwidth}{lX}
    \toprule
    ID  & Short label (positive only) \\ \midrule
    $Q_1$  & Intended frequency of use \\
    $Q_2$  & System simplicity \\
    $Q_3$  & Ease of use \\
    $Q_4$  & User independence \\
    $Q_5$  & Functions well integrated \\
    $Q_6$  & Design consistency \\
    $Q_7$  & Learnability \\
    $Q_8$  & Navigation smoothness \\
    $Q_9$  & User confidence \\
    $Q_{10}$ & Low learning effort \\ \bottomrule
\end{tabularx}
 \end{table}

To quantify {\Builder}'s \emph{usability}~\ref{rq1-usab}, we employed the system usability scale (SUS)~\cite{brookeSUSQuickDirty1996}. SUS is a robust, industry-standard metric in usability research, applied across diverse domains~\cite{broekhuisAssessingUsabilityEHealth2019,vlachogianniPerceivedUsabilityEvaluation2022, ferrari2021systematic}, including other MDE tools (e.g., \textsc{Extremo}~\cite{moraseguraSosymExtremoEval2024}). 
It consists of 10 items rated on a five-point Likert scale with alternating positive and negative statements to mitigate response bias. Final scores are normalized to a scale ranging from 0 (the worst imaginable) to 100 (the best imaginable). For this study, the SUS phrasing was adapted to explicitly reference {\Builder}, while preserving the standardized scoring procedure. We also associate a short label to each item, which we will refer to subsequently, available in Table~\ref{tab:sus-builder}. The complete version can be found in Table~\ref{tab:sus_builder_full}.

To assess the perceived adequacy~\ref{rq2-feat}, we utilized a dedicated questionnaire where participants rated key interface components on a five-point scale (1: Very Demanding, 5: Very Intuitive). Specifically, we evaluated:
\begin{itemize}
    \item \emph{pipeline modelization} using FDF concepts on the {\Builder} canvas~\ref{ui-canvas},
    \item \emph{box parametrization} via the parameter editor~\ref{ui-param},
    \item \emph{result visualization} in the integrated chart viewer~\ref{ui-chart},
    \item \emph{warning utility}, specifically the interpretation of feedback messages (e.g., Fig.~\ref{fig:fdf-type-warning}).
\end{itemize}

We also analyzed the open-ended qualitative responses and identified recurring themes such as positive mentions, usability barriers, and improvement suggestions.

\subsection{Participant Profiles}

\begin{figure*}[t!]
    \centering
    \begin{subfigure}[b]{0.48\textwidth}    
        \centering
        \includegraphics[width=\textwidth]{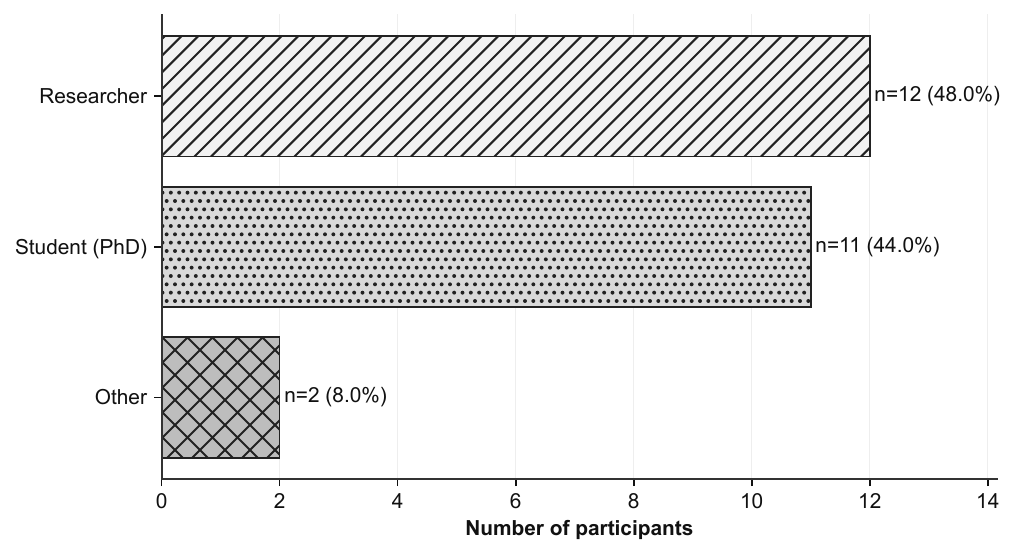}
        \caption{Breakdown of study participants by their role}
        \label{fig:role_distribution}
    \end{subfigure}
    \hfill
    \begin{subfigure}[b]{0.48\textwidth}
        \centering
        \includegraphics[width=\textwidth]{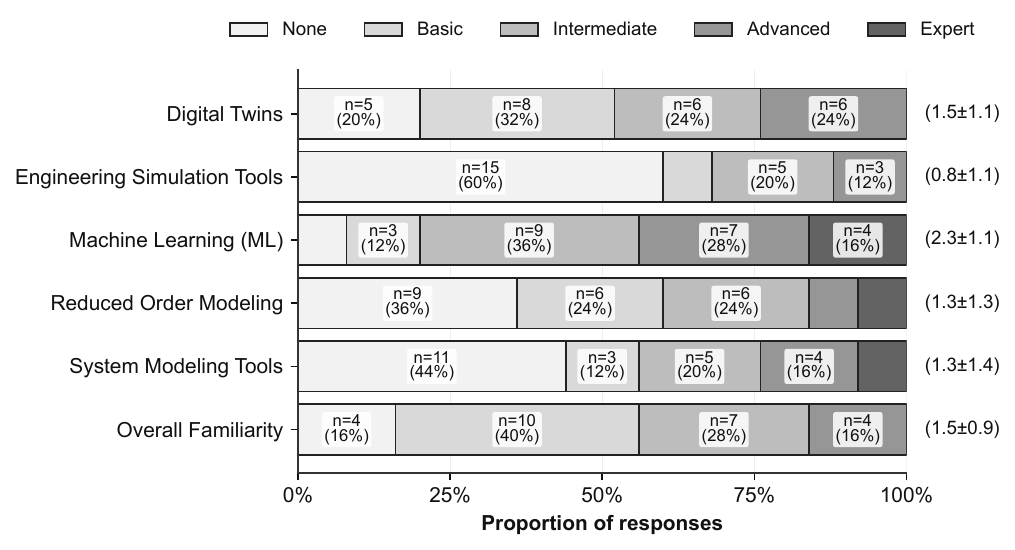}
        \caption{Familiarity with key concepts and tools related to {\Builder}}
        \label{fig:familiarity_ratings}
    \end{subfigure}
    \vfill
    \begin{subfigure}[b]{0.48\textwidth}
        \centering
        \includegraphics[width=\textwidth]{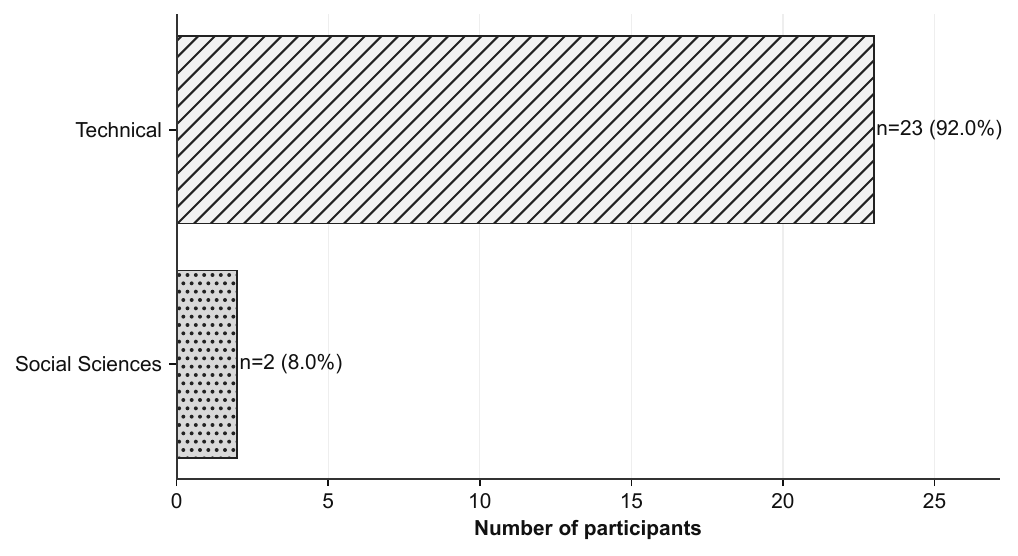}
        \caption{Primary area of expertise of study participants}
        \label{fig:expertise_distribution}
    \end{subfigure}
    \caption{Participant demographics in the {\Builder} user study}
    \label{fig:user_study_demographics}
\end{figure*}

A total of 26 participants were recruited across the two sessions, $S_1$ -- $S_{26}$. Following standard data quality screening, the participant $S_{14}$ was excluded for extreme response speeding and invariant response (straight-lining)~\cite{leinerTooFastToo2019}, resulting in a final sample of $N=25$. As shown in Fig.~\ref{fig:role_distribution}, the final cohort primarily consisted of researchers ($48.0\%$, $n=12$) and PhD students ($44\%$, $n=11$), with the remaining $8.0$\% ($n=2$) having other roles.

We assessed prior expertise across five domains central to the FDF framework (Fig.~\ref{fig:familiarity_ratings}). Participants were most familiar with ML ($2.3\pm1.1$) and least with engineering simulation tools ($0.8\pm1.1$). Intermediate familiarity was reported for digital twins, reduced-order modeling, and system modeling tools. 
To characterize the cohort's background, we calculated an \emph{overall familiarity score} as the per-respondent mean across all categories, mapped to the nearest integer on a 0 (None) to 4 (Expert) scale.

We also surveyed participants on their primary area of expertise (Fig.~\ref{fig:expertise_distribution}). Most participants ($92.0\%$, $n=23$) declared that their expertise is in a technical field (AI/ML, systems engineering, among others), while the remaining $8.0$\% ($n=2$) have expertise in social sciences.

\subsection{Results} \label{sec:user-study-results}

This section reports the results of our user study, focusing on the evaluation of our two research questions (RQs). 

\subsubsection[(RQ1) Perceived usability of {\Builder} and FDF]{\ref{rq1-usab} Perceived usability of {\Builder} and FDF}

\begin{figure}[h!]
    \centering
    \includegraphics[height=0.25\textheight]{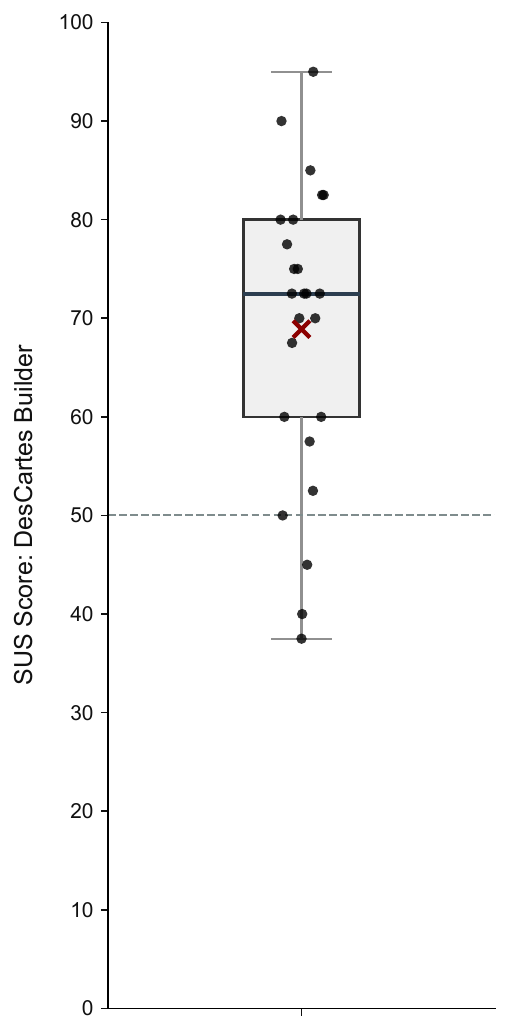}
    \caption{Box plot with the normalized SUS score (0 - 100 scale, higher = better). Black line: median (72.5); dark red cross ({\color{darkred}$\times$}): mean score (68.9); dashed line: unnaceptable score (50)}
    \label{fig:sus_boxplot}
\end{figure}

Our analysis indicates that {\Builder} achieves a \emph{good} usability rating~\cite{bangorEmpiricalEvaluationSystem2008}. 
As illustrated in Fig.~\ref{fig:sus_boxplot}, the median SUS score is \emph{72.5}, while the mean SUS score is $\emph{68.9}\pm15.2$ (95\% CI: $[62.6,75.2]$; range: 37.5 -- 95).
The median exceeding the mean indicates a generally favorable experience, with the mean skewed by a few low-scoring outliers whose low familiarity scores placed them outside our primary target audience.
Notably, only 16\% of participants rated the tool below the "unacceptable" threshold of 50~\cite{bangorEmpiricalEvaluationSystem2008}.

The statistical approach was validated by a Shapiro-Wilk test, which shows that the data do not significantly deviate from a normal distribution: $W=0.95, p=0.29$, where $W=1$ represents a perfect normal distribution. Consequently, the 95\% CI (confidence interval) reported was obtained using a T-test.

\paragraph{Comparative context}

\begin{table}[h]
    \centering
    \caption{Mean SUS Scores for selected tools}
    \label{tab:sus_scores}
    \begin{tabular}{lll}
        \toprule
        \textbf{Tool}    & \textbf{SUS Score} & \textbf{Reference} \\
        \midrule
        Simulink         & 76.4              & \cite{ferrari2021systematic}             \\
        UPPAAL           & 61.7              & \cite{ferrari2021systematic}              \\
        Excel            & 56.5              & \cite{kortumUsabilityRatingsEveryday2013}               \\
        \textsc{Extremo} & 70.0              & \cite{moraseguraSosymExtremoEval2024}               \\
        \rowcolor{gray!30}\Builder & 68.9    &  \cite{decontoFunction+DataFlowFramework2024,decontoDesCartesBuilderTool2025}             \\ 
        \bottomrule
    \end{tabular}
\end{table}

To contextualize these findings, Table~\ref{tab:sus_scores} compares {\Builder}'s mean SUS score against established tools. We report mean scores only, as median scores are not always provided for other tools. 
Despite being a specialized research prototype at an early stage of development, {\Builder}'s score (68.9) exceeds that of some mature tools like Excel (56.5) and UPPAAL (61.67).

\paragraph{Item-level analysis}

\begin{figure}[t!]
    \centering
    \includegraphics[width=\columnwidth]{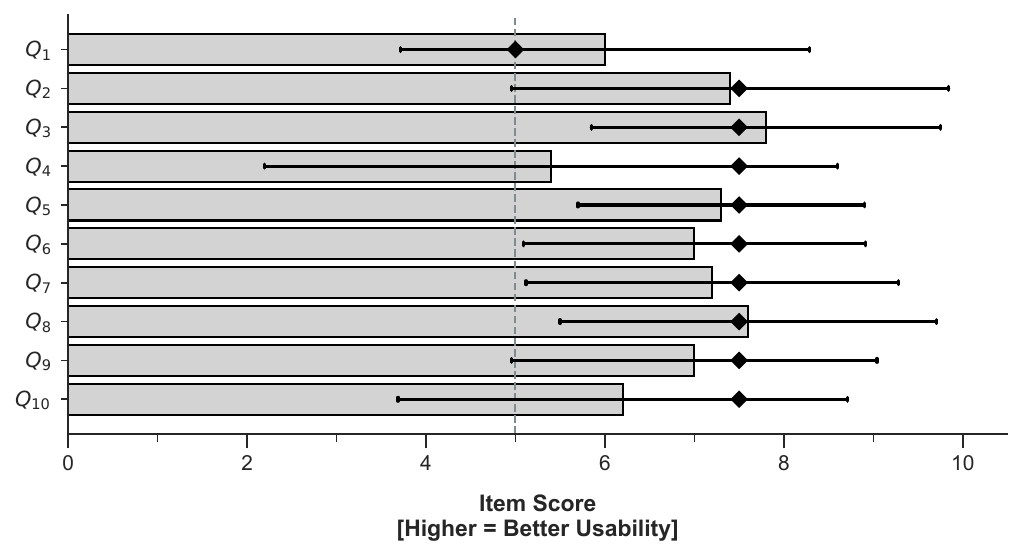}
    \caption{Normalized SUS item scores (0-10 scale, higher = better). Error bars: standard deviation; diamonds: medians; dashed line: unacceptable score (5.0)}
    \label{fig:sus_scores}
\end{figure}

An item-level analysis of the SUS scores (Fig.~\ref{fig:sus_scores}) reveals that, while all items remain acceptable ($>5.0$), questions $Q_1$ (Intended frequency of use), $Q_4$ (User independence), and $Q_{10}$ (Low learning effort) emerged as the primary factors lowering the overall score. 
Notably, $Q_4$ and $Q_{10}$ constitute learnable SUS sub-score~\cite{lewisFactorStructureSystem2009}, which for {\Builder} stands at \emph{58.0}, significantly lower than its global mean 68.9. This suggests that some participants faced a steep learning curve.

\paragraph{Impact of technical familiarity}

\begin{table}[h]
\centering
\small 
\caption{SUS scores for subgroups}
\label{tab:sensitivity_compact}
\begin{tabular}{lrr}
\toprule
\textbf{Group ($n$)} & \textbf{Mean (SD)} & \textbf{95\% CI} \\ 
\midrule
Technical (15)       & 72.8 (12.1)        & [66.1, 79.5]     \\
Low-technical (10)   & 63.0 (17.9)        & [50.2, 75.8]     \\
\midrule
\textbf{Total (25)}  & \textbf{68.9 (15.2)} & \textbf{[62.6, 75.2]} \\
\bottomrule
\multicolumn{3}{l}{\footnotesize \textit{Note:} Welch's $t(14.4)=1.52, p=0.15, d=0.67$.}
\end{tabular}
\end{table}

To evaluate the impact of technical familiarity (Fig.~\ref{fig:familiarity_ratings}), we conducted an exploratory sensitivity analysis (Table~\ref{tab:sensitivity_compact}). By excluding the low-technical subgroup ($n=10$, bottom quartile of overall familiarity), the mean SUS score for the remaining technical participants ($n=15$) increased to $72.8\pm12.1$ (95\% CI: $[66.1, 79.5]$). The moderate effect size (Cohen's $d=0.67$) suggests that {\Builder} is most effective for its intended audience of \emph{domain experts}, whereas very low technical proficiency can present different barriers to adoption. 

\subsubsection[(RQ2 Feature adequacy of {\Builder} and FDF]{~\ref{rq2-feat} Feature adequacy of {\Builder} and FDF}

\begin{figure}[t!]
    \centering
    \includegraphics[width=\columnwidth]{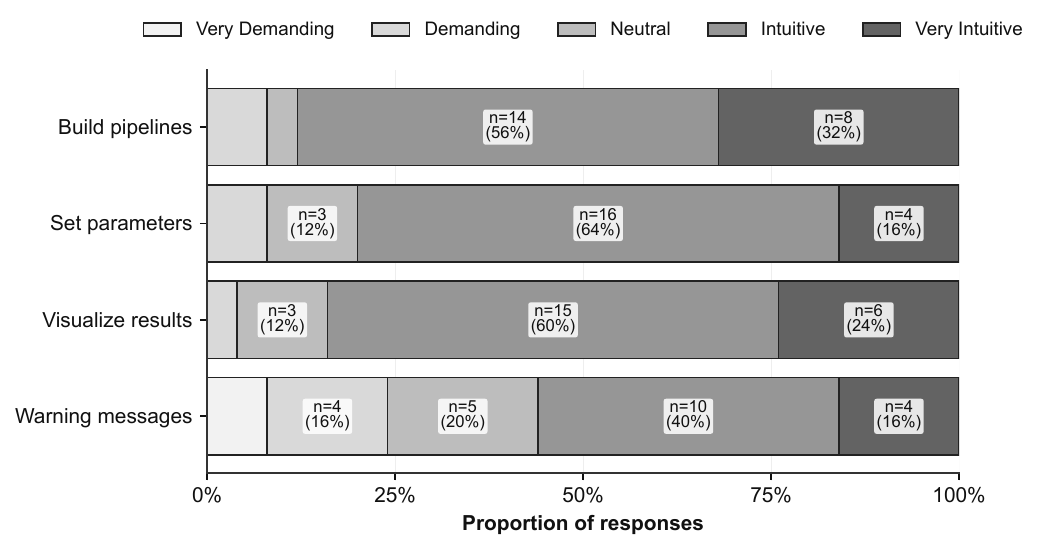}
    \caption{Participant feedback on specific {\Builder} features}
    \label{fig:user_friendly_distribution}
\end{figure}

Feature-specific ratings were uniformly positive for {\Builder} (Fig.~\ref{fig:user_friendly_distribution}): \emph{89\%} of participants rated the canvas~\ref{ui-canvas} intuitive or very intuitive; \emph{81\%} found parameter setting with the parameter editor~\ref{ui-param} straightforward, and \emph{85\%} were satisfied with the results visualization using the chart viewer~\ref{ui-chart}. 
The overall mean feature rating was $3.9 \pm 0.6$ on a 5-point scale (95\% CI: $[3.0, 4.1]$). Normality was verified with a Shapiro-Wilk test ($W = 0.93, p = 0.07$), justifying the use of a T-test for the confidence interval.

However, feedback regarding the warning messages was more polarized; only 57\% of participants considered the warning messages to be "friendly".

\subsection{Discussion}

We now present some points of discussion based on the results from Sect.~\ref{sec:user-study-results}: an analysis of low-scoring participants, key positive feedback, areas where there is room for improvement, and a discussion on the validity of our results. 

\subsubsection{Analysis on low-scoring participants}

As shown in Fig.~\ref{fig:sus_boxplot}, only 4 participants (16\%) reported low usability (SUS$\le$ 50). 

Two of these outliers ($S_{20}$, SUS=37.5; $S_{22}$, SUS=40.0) were those with primary expertise in social sciences (Fig.~\ref{fig:expertise_distribution}). This indicates that \Builder's reliance on technical abstractions (e.g., block diagrams) may impose a significant cognitive load on them~\cite{yusufInteractionPatternsBlockbased2024}. Qualitative feedback supports this interpretation: $S_{20}$ states challenges to decide "where to click" and $S_{22}$ explicitly mentioned a need for "technical guidance". To mitigate this, future versions should include optional tutorials: technical users shall have the option to quickly skip it since the extra guidance can be counterproductive for them~\cite{kalyugaExpertiseReversalEffect2009}.

The other two outliers ($S_{10}$ and $S_{24}$) appear to experience a \emph{"paradigm clash"}.
$S_{24}$, an ML expert, noted that the "modelling logic differs from 'traditional' approaches" and requested "more flexibility", suggesting that the FDF paradigm may feel restrictive to users accustomed to code-based, imperative workflows. This aligns with known MDE trade-offs where increased abstraction can lead to a loss of granular control among experts~\cite{whittleStatePracticeModelDriven2014}. 
Crucially, the FDF framework is inherently flexible, permitting arbitrary logic within its boxes. This restriction is therefore specific to the current \Builder implementation. To address this, future versions of \Builder will allow users to define and extend FDF boxes with custom code, similarly to, e.g., KNIME's Python Script node\footurl{https://hub.knime.com/knime/extensions/org.knime.features.python3.scripting/latest/org.knime.python3.scripting.nodes2.script.PythonScriptNodeFactory}.
On the other hand, $S_{10}$ noted that the "connection between blocks" was confusing and suggested "combining \Trainer and \Processor" boxes to reduce complexity. Note we had previously identified this requirement, and its support is included as the \Module box in the extension proposed in Sect.~\ref{sec:hfdf}.

\subsubsection{Positive feedback}

We observe that users highly valued {\Builder}'s \emph{visual and intuitive graph interface}. For example, $S_1$ highlighted that the "plug-and-play blocks" make the system easier to use, and $S_8$ appreciated that there is "no need to write code".

The FDF modeling paradigm was also well-received, though feedback suggests a slight initial learning curve. For instance, a participant noted that "the modeling part was good," adding that "once you understand the red line is a function flow, it also feels very intuitive". This learning curve may be influenced by a user's background; those familiar with Model-Driven Engineering (MDE) tools appeared to adapt more quickly. Supporting this, another participant explicitly observed that the interface "felt similar to using Simulink."

Finally, participants valued {\Builder} as an \emph{integrated platform}. A user praised it for having "one platform for various supervised ML algorithms", a feature that directly enables \emph{rapid experimentation}. As another participant stated, this enables users to "change the model quickly and run the experiments" without switching contexts.

The workshop also indicates that the tool already has a high degree of cross-platform stability. All users were able to install the tool on their own machine (macOS, Linux, or Windows) using the provided installers and guidance. This indicates that {\Builder} already achieves a very good level of quality and technical maturity in its first release. 

\subsubsection{Key areas for improvement}

The survey also provides insight into two main types of challenges that users have faced in  {\Builder}: lack of tooltips and built-in documentation, and need for further diagnostics and runtime inspections.

\paragraph{Tooltips and built-in documentation}

The SUS item analysis (Fig.~\ref{fig:sus_scores}) and the learnable SUS sub-score~\cite{lewisFactorStructureSystem2009} of 58.0, significantly lower than the global mean at 68.9, suggest an initial adoption barrier due to a steep learning curve for some users. 
Qualitative feedback confirms that documentation is an important challenge, particularly as the tool introduces unfamiliar concepts, with $36\%$ of users providing documentation-related feedback\footnote {i.e., including keywords "document", "tutorial", "guide", "tooltip", "tips"}. For instance, participants identified a "lack of understanding [of] the role of different modules" and recommended "pop-up tutorials", or "tooltips for the different components". 

We conclude that users expect built-in guidance in the tool (e.g., tooltips, dedicated onboarding) as per what is now an established practice in the industry~\cite{grossmanToolClipsInvestigationContextual2010,huckoYesElfPersonalizedOnboarding2019}. This is a key direction we aim to improve {\Builder} in future releases to turn it from a high-quality research prototype into a production-ready tool. 

\paragraph{Diagnostics and runtime inspection}

While the tool's core modeling features achieved high satisfaction, the polarized feedback on warning messages (57\% "friendly") indicates that this is a key area for improvement in the tool. These findings are corroborated by the qualitative feedback, with 24\% of users providing diagnostics-related feedback\footnote{i.e., including keywords  "warning", "error", "diagnostic", "runtime", "inspect", "console", "log"}. 

Participants would like to move beyond final results by, e.g., "inspecting data flowing through the pipeline" and viewing metadata like "dimensions or data types." Participants also stated that they encountered errors or warnings where they "don't know what the causes are" and suggested that the tool should provide specific "suggestions to resolve" detected issues. We view these requests as a positive sign of user engagement, suggesting that participants would like to move to more sophisticated debugging and optimization. Thus, we will similarly consider this a high-priority improvement for {\Builder}. 

\subsection{Discussion on validity}

We now discuss internal validity and external validity of our study, based on~\cite{wohlinExperimentationSoftwareEngineering2024}.

\emph{Internal validity.}
Supervision of the workshop by the study authors could have encouraged more favorable feedback. To mitigate this, we explicitly emphasized anonymity and the importance of critical feedback~\cite{singerSoftwareEngineeringData2008}. 
Our discussion and qualitative feedback highlight that users provided important and constructive feedback on the challenges they faced and did not hesitate to identify shortcomings in the tool. We therefore believe this threat was adequately mitigated.

\emph{External validity.} 
While the cohort included students, we believe that their background in ML and engineering makes them suitable proxies for early-career practitioners in the DT domain~\cite{salmanAreStudentsRepresentatives2015, falessiEmpiricalSoftwareEngineering2018}. Additionally, all our students were at the PhD level, representing more advanced technical expertise. We also mitigated this threat by including researchers who have deeper technical expertise that may more closely represent the domain experts targeted by the tool. 
Regarding task scope, the two-hour session may have limited participants to simpler scenarios, excluding complex, multi-domain DT pipelines and long-term maintainability. 
Nevertheless, this duration is consistent with (and exceeds) the one-hour sessions common in usability studies~\cite{sauroSUSbook}, and was deliberately chosen to accommodate the added complexity of {\Builder} while balancing participant fatigue and task representativeness.

\section{The H-FDF extension}
\label{sec:hfdf}

The original FDF has two main limitations. 
First, its strictly acyclic architecture cannot represent iterative processes unless hard-coded inside a function, which contradicts the modular philosophy of a DSL approach. 
Second, FDF provides no native support for subcomponents (or modules). 
Although this does not limit FDF's expressive power, it makes composing or reusing the same construct tedious. Our user study (Sect.~\ref{sec:empirical-study}) evidenced this. Participant ($S_{10}$) suggested combining the \Trainer and associated \Processor box to "reduce the complexity in designing the pipelines", which is easily implemented with a module.  

We address both limitations by explicitly defining a \emph{module} construct. This enables compositional specification~\cite{ekerTamingHeterogeneityPtolemy2003} and creates a \emph{hierarchical pipeline}: higher-level modules can call lower-level ones (but not the other way around, keeping the workflow strictly unidirectional). 
Crucially, while our modules can be locally (internally) inductive, the pipeline remains \emph{globally acyclic} when called externally. This differs from KNIME, where iterations are defined with special loop-start/loop-end nodes (globally, KNIME pipelines are also acyclic). Further, while modules exist in KNIME (called components or metanodes), they are always \emph{non-inductive}.

Our proposal introduces a unique type of \emph{iterative modules} that handles the variability of what to iterate on, specifically the stopping criterion and the data or functions to refine iteratively. 
This is achieved through explicit flows between outputs and inputs of the module (in the module definition, not at the higher-level when calling the module). This is possible because, at design time, types are \emph{implicit} (Sect.~\ref{sec:implicit-typing}) and need to be resolved only at runtime. We call this extension \emph{Hierarchical Function+Data Flow (H-FDF)}.

\subsection{Extended FDF} \label{sec:extended-fdf}

\begin{figure}[t!]
    \centering
    \includegraphics[width=\columnwidth]{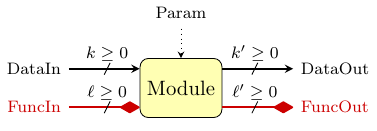}
    \caption{Visual syntax for the \Module box. The \Module box encapsulates a subpipeline executed upon its invocation from a higher-level \Module}
    \label{fig:hier-fdf-boxes}
\end{figure}

Before we define the full H-FDF, we start by describing the \emph{syntax and semantics of Extended FDF (E-FDF)}, which executes one subpipeline, possibly iteratively. 

On top of FDF, E-FDF defines a fourth box class, namely \Module, in addition to the three box classes from standard FDF (\Processor, \Coder, and \Trainer). 
The \Module box is depicted as a rounded, pale-yellow rectangle (Fig.~\ref{fig:hier-fdf-boxes}). It has $k \geq 0$ \IDP s to receive data to be processed and $\ell \geq 0$ \IFP s to receive functions used in the hierarchical pipeline. It returns $k' \geq 0$ \ODP s produced by the subpipeline and $\ell' \geq 0$ {\OFP} functions produced/refined. Importantly, in E-FDF, each \Module encapsulates a (possibly iterative) subpipeline executed internally (see Sect.~\ref{sec:hfdf-syntax}).

In addition, E-FDF supports \emph{function initialization} in \Trainer and \Coder boxes. While in FDF, the \Trainer and \Coder boxes have \emph{zero} input function ports, in E-FDF we allow \emph{zero or one} input function ports. If a function is provided, it serves to initialize the model in these boxes, rather than standard random initialization in FDF. This is crucial for iterative function learning.

\subsubsection{Formal E-FDF syntax}

We define the E-FDF syntax by extending the FDF syntax~\cite{decontoFunction+DataFlowFramework2024} with a \Module box class and support for iterations. Formally, an E-FDF pipeline, denoted $\FDpipe$, is the tuple:
\begin{multline*}
\FDpipe=(\FDboxes', \boxclass, \FDports, \portclass, \portbox,\src,\\ \param, [topin, topout, stop, \feedback]),
\end{multline*}
where brackets $[\cdot]$ indicate new optional elements: if none of these optional elements are present, E-FDF reduces to the original FDF. In addition, E-FDF extends the mappings $\boxclass$ and $\param$ as required: $\boxclass$ includes  the \Module box, and $\param$ supports a Boolean expression that specifies the iteration stop condition. The remaining mandatory elements are defined identically to FDF~\cite{decontoFunction+DataFlowFramework2024}:
\begin{itemize}
    \item Let $\FDboxes =  \{\FDbox_1, \cdots , \FDbox_{n}\}$ be the set of user-defined boxes.
    We denote $\FDboxes' = \FDboxes \sqcup \{\DataIO,\FuncIO\}$, including the pipeline data and function dependencies.
    \item $\boxclass: \FDboxes \rightarrow \{\Processor, \Coder,\allowbreak \Trainer,\Module\}$ defines the class of each box.
    \item $\FDports = \FDports^I \sqcup \FDports^O = \{1, \cdots, m\}$ is the ordered set of natural numbers up to $m$, the number of ports. This set is partitioned into the sets of input ports $\FDports^I$ and the set of output ports $\FDports^O$.
    \item $\portclass: \FDports \rightarrow \{\text{Data, Function}\}$ provides the class (Data or Function) of each port.
    \item $\portbox: \FDports \rightarrow \FDboxes'$ is a function associating each port with the box it belongs to.  
    \item $\src: \FDports^I \rightarrow \FDports^O$ is a function mapping every input port to its corresponding source output port (i.e., the one providing its data/function). $\src$ is such that an input port $\FDport$ and the output port it is connected to have the same class: 
    $\forall \FDport \in \FDports^I, \portclass(\FDport) = \portclass(\src(\FDport))$.
    \item $\param: \FDboxes \rightarrow \mathsf{BoolExpr} \cup \Library \, \cup \,  \mathbb{N} \times \Library$ provides the parameters of each box. $\Library$ is a library of predefined functions, $\mathbb{N}$ denotes the number of $X$ input to \Trainer and $\mathsf{BoolExpr}$ is the set of Boolean expressions (e.g., $a > 0$). 
\end{itemize}

The optional elements $[topin, topout, stop,\allowbreak \feedback]$ are defined as: 
\begin{itemize}   
    \item $topin,topout \geq 1$ define the boundary ports of the pipeline. The \emph{first} $topin$ ports are the top-level inputs for this pipeline, such that for each port $1 \leq \FDport \leq topin$, we have $\FDport \in \FDports^O$. 
    Similarly, the \emph{last} $topout$ ports are the top-level outputs for this pipeline, such that for each port $m-topout+1 \leq \FDport \leq m$, we have 
    $\FDport \in \FDports^I$. The port $\FDport$ has $\portbox(\FDport)=\DataIO$ if $\portclass(\FDport)=\text{Data}$ and $\portbox(\FDport)=\FuncIO$ if $\portclass(\FDport)=\text{Function}$.
    
    \item $stop$ is a \Processor box with a unique input and a Boolean stopping condition, provided as a parameter. 
    The iteration stops whenever the unique input satisfies the Boolean condition. If there is no $stop$ box, the condition is considered true, and the \Module stops after a single iteration. 

    \item $\feedback :  \{m-topout+1,\ldots, m\} \nrightarrow \{1,\ldots, topin\} $
    is a partial function associating some of the top-level output ports with some top-level input ports, specifying which data/function are iterated over.
    $\feedback$ is such that the $topout$ ports in the domain of $\feedback$ are associated with $topin$ ports sharing the same class:
    $\forall \FDport \in \dom(\feedback), \portclass(\FDport) = \portclass(\feedback(\FDport))$. The ports listed in $\feedback$ are said to be \emph{recursive} ports and are depicted as a square ($\blacksquare$) in the \Module boundary, while the remaining ports are depicted as a circle ($\bullet$).
    
\end{itemize}

An E-FDF pipeline is said to be {\em non-inductive} when the optional elements $stop$ and $\feedback$ are omitted, resulting in a single iteration. When included, the pipeline is \emph{inductive}, supporting recurrent data and function flows (defined by $\feedback$) and iterating until a stopping condition specified with the $stop$ \Processor box.

\subsubsection{E-FDF semantics} \label{sec:efdf-semantics}

\begin{figure*}[t!]
    \centering
    \begin{subfigure}[b]{0.45\textwidth}
        \centering
        \includegraphics[width=1.2\textwidth]{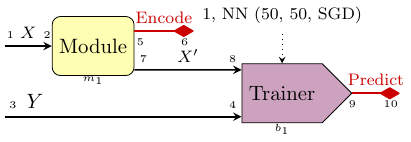}
        \caption{Topmost subpipeline $\FDpipe_1$ with a Module box $\FDmod_1$}
        \label{fig:outer_pipeline}
    \end{subfigure}
    \hfill
    \begin{subfigure}[b]{0.45\textwidth}
        \centering
        \includegraphics[width=0.7\textwidth]{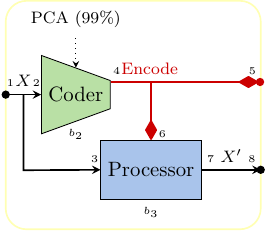}
        \caption{Non-inductive subpipeline $\FDpipe_2=\hier(\FDmod_1)$}
        \label{fig:inner_pipeline}
    \end{subfigure}
    \caption{Example of a H-FDF pipeline $\FDpipe_1,\FDpipe_2,\hier$}
    \label{fig:hierarchical_pipeline_example}
\end{figure*}

The E-FDF semantics extends the FDF semantics, defining how a \Module box is executed. Running a \Module box with $k$ {\IDP}, $\ell$ {\IFP}, $k'$ {\ODP} and $\ell'$ {\OFP} proceeds as follows: 
\begin{enumerate}
    \item \emph{Initialization:} Load the internal data and functions from the \DataIO and \FuncIO ports, and load the top-level inputs 
    ($1 \leq \FDport \leq topin$).
    
    \item \emph{Execution:} Run the E-FDF pipeline as per the partial order of the E-FDF graph, which is defined as the FDF graph (Sect.~\ref{sec:fdf-pipe-ports}). The $\feedback$ edges are not considered, the graph remaining acyclic. \Module boxes execute as fixed functions on their inputs until they provide their outputs.
    
    \item \emph{Stop condition:} Evaluate the condition $cond=\param(stop)$ on the unique input of the box $stop$:
    \begin{itemize}
        \item \emph{Termination:} If $cond$ is true, the workflow stops: the \DataIO and \FuncIO input ports provide their final values to the top-level outputs ($m-topout+1 \leq \FDport \leq m$).
        \item \emph{Iteration:} If $cond$ is false, the top-level inputs are updated for the next iteration: all ports $1 \leq \FDport \leq topin$ such that $\FDport' \geq m-topout+1$ with $\feedback(\FDport')=\FDport$ are loaded from $\FDport'$. All other ports retain their values from the previous iteration 
        and the pipeline repeats from Step 2.
    \end{itemize}
\end{enumerate}

\subsubsection{E-FDF implicit typing}

E-FDF implicit typing follows FDF's implicit typing. \Module boxes are considered as fixed functions in E-FDF. Hence, the user (here H-FDF, see Section \ref{sec.hfdftype}) provides an implicit typing scheme defining how each output is related to the inputs of the box, or whether an output is a fresh type.

We now turn to H-FDF.

\subsection{Syntax of H-FDF}
\label{sec:hfdf-syntax}

A hierarchical FDF (H-FDF) pipeline is formally defined as a finite sequence \FDhierpipe\ of $n_{\FDpipe}$ subpipelines together with a mapping $\hier$, where: 

\begin{itemize}
    \item $\FDhierpipe = \FDpipe_1, \ldots, \FDpipe_{n_{\FDpipe}}$,
    \item each subpipeline $\FDpipe_i$ is an E-FDF pipeline,
    \item $\FDpipe_{1}$, called the {\em topmost} pipeline, does not feature $topin, topout, stop, \feedback$,
    \item $\FDpipe_{n_{\FDpipe}}$ has no \Module box and is {\em non-inductive},
    \item $\hier$ associates each \Module box of each $\FDpipe_i$ 
    with a subpipeline $P_{j>i}$. It is required that the number of input and output ports of each \Module $\FDmod$ matches the number of $topin$ and $topout$ ports of $\hier(\FDmod)$ respectively, as well as their types (data or function).
\end{itemize}

\subsubsection{Annotated Example}

We now review a simple H-FDF pipeline (Fig.~\ref{fig:hierarchical_pipeline_example}): 
$P_1$ uses a module $\FDmod_1$ calling pipeline $P_2$ (Fig.~\ref{fig:outer_pipeline}). The subpipeline $P_2$ computes the {\tt Encode} function from a dataset, applies it to the dataset, and outputs both (Fig.~\ref{fig:inner_pipeline}). 

Here is the formal syntax for this H-FDF pipeline 
$\FDhierpipe = \{\FDpipe_1, \FDpipe_2\}$, with 
$\hier(\FDmod_1)=\FDpipe_2$. 
The top-most pipeline $\FDpipe_1$ is defined as: 
\begin{itemize}
    \item $\FDboxes = \{\FDbox_1, \FDmod_1\}$, with 
    $\boxclass(\FDbox_1) =$ \Trainer, $\boxclass(\FDmod_1) =$ \Module.
    \item $\param(\FDbox_1) =$ 1, NN (50, 50, SGD), meaning a \Trainer with 1 $X$ input and the remaining port as $Y$, learning a neural network (NN) with 2 layers of 50 nodes each using stochastic gradient descent (SGD).
    \item $\FDports = \FDports^I \sqcup \FDports^O$, with $\FDports^I = \{2,4,6,8,10\}$ and $\FDports^O = \{1,3,5,7,9\}$.
    \item $\portclass(\FDport) = \\\begin{cases}
        \text{(Data)},     & \text{if } \FDport \in \{1,2,3,4,7,8\} \\
        \text{(Function)}, & \text{if } \FDport \in \{5,6,9,10\}\\
    \end{cases}$      
    \item $\portbox(\FDport) = \begin{cases}
        \DataIO,   & \text{if } \FDport \in \{1, 3\} \\
        \FuncIO,  & \text{if } \FDport \in \{6, 10\} \\
        \FDmod_1, & \text{if } \FDport \in \{2, 5, 7\} \\
        \FDbox_1, & \text{if } \FDport \in \{4, 8, 9\} \\
    \end{cases}$
    \item $\src(2) = 1;
           \src(4) = 3;
           \src(6) = 5;
           \src(8) = 7;
           \src(10) = 9.$
    \item $stop = \bot$, $\feedback=\emptyset$.
    \item $topin = 0$, $topout = 0$. 
\end{itemize}

The sub-pipeline $\FDpipe_2$ is given by:
\begin{itemize}
    \item $\FDboxes = \{\FDbox_2, \FDbox_3\}$.
    \item $\boxclass(\FDbox_2) =$ \Coder; 
          $\boxclass(\FDbox_3) =$ \Processor.
    \item $\param(\FDbox_2) =$ PCA($99\%$).
    \item $\FDports = \FDports^I \sqcup \FDports^O$ with 
    $\FDports^I = \{2,3,5,6,8\}$ and $\FDports^O = \{1,4,7\}$.
    \item $\portclass(\FDport) = \\\begin{cases}
        \text{(Data)},     & \text{if } \FDport \in \{1,2,3,7,8\} \\
        \text{(Function)}, & \text{if } \FDport \in \{4,5,6\}\\
    \end{cases}$      
    \item $\portbox(\FDport) = \begin{cases}
        \DataIO, & \text{if } \FDport \in \{1,8\} \\
        \FuncIO, & \text{if } \FDport \in \{5\} \\
        \FDbox_2, & \text{if } \FDport \in \{2,4\} \\
        \FDbox_3, & \text{if } \FDport \in \{3,6,7\}
    \end{cases}$
    \item $\src(2) = 1;
           \src(3) = 1;
           \src(5) = 4;
           \src(6) = 4;
           \src(8) = 7;
           $
    \item $stop = \bot$, $\feedback=\emptyset$.
    \item $topin = 1$ with $1 \in \FDports^O$, and $\portbox(1) = \DataIO$ since  $\portclass(1) = \text{Data}$.
    \item $topout = 2$ with $\{5,8\} \subset \FDports^I$, where 
    $\portbox(5) = \FuncIO$ since $\portclass(5)=\text{Function}$, and 
    $\portbox(8) = \DataIO$ since $\portclass(8)=\text{Data}$.
\end{itemize}

Ports are matched based on the natural ordering:
\DataIO port $1$ of $\FDpipe_2$ is matched with {\IDP} 2 of $\FDmod_1$.
\FuncIO port $5$ of $\FDpipe_2$ is matched with
{\OFP} 5 of $\FDmod_1$, while \DataIO port $8$ of $\FDpipe_2$ is matched with {\ODP} 7 of $\FDmod_1$.

\subsection{Formal Semantics}

H-FDF retains the key components of FDF formal semantics. Instead of a single graph associated with the FDF pipeline, an H-FDF pipeline $\FDhierpipe = \{\FDpipe_1, \ldots, \FDpipe_{n_{\FDpipe}} \}$ has one directed graph $G_i=G(\FDpipe_i)$ for each sub-pipeline. The same graph definition $G(\FDpipe_i) = (\FDports, \FDedges)$ is used for H-FDF and FDF, as in Sect.~\ref{sec:fdf-pipe-ports}.

The execution semantics are extended from FDF, executing each 
E-FDF box $\FDbox$ according to its class. 
For a \Module box $\FDmod$, the pipeline $\hier(\FDmod)$ is executed as explained in Sect.~\ref{sec:efdf-semantics}, outputting the data and function in the {\OP} of $\FDmod$.

\subsection{Implicit Typing Resolution}
\label{sec.hfdftype}

H-FDF resolves the implicit typing of E-FDF for the \Module box class in the implicit type inference (Phase III).

A \Module $\FDmod$ with $topout$ top-level outputs determines these types by traversing the module's internal subpipeline $\hier(\FDmod)$. Concretely, the type of each $topout$ output port is the type propagated to these ports during propagation (Phase II).
An \emph{inconsistent input type} warning is raised whenever $\exists \FDport \in \dom(\feedback)$ such that $\type(\FDport) \ne \type(\feedback(\FDport))$: a $topout$ port can only be connected, via $\feedback$, with $topin$ ports that have the same implicit type.

\subsection{Case study: Dual Training}
\label{sec:case}

\begin{figure*}[t]
    \centering
    \begin{subfigure}[b]{0.48\textwidth}
        \centering
        \includegraphics[width=\linewidth]{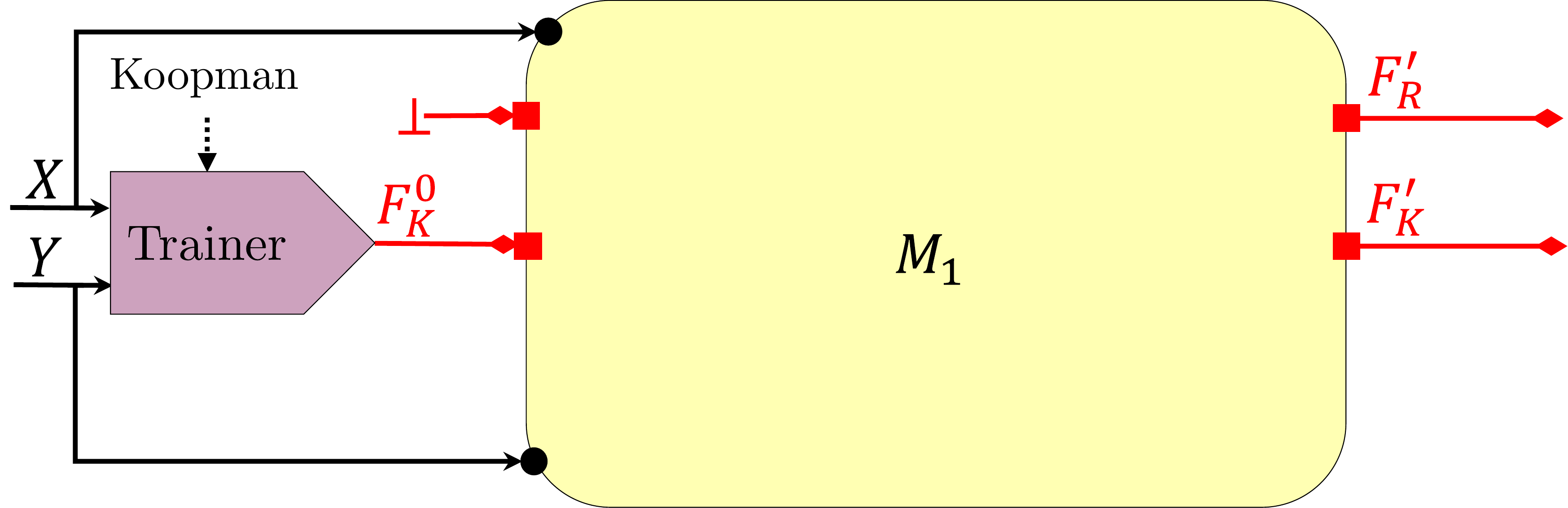}
        \caption{Topmost subpipeline $\FDpipe_1$ with $\hier(\FDmod_1)=\FDpipe_2$}
        \label{fig:model-comb-overview}
    \end{subfigure}
    \hfill
    \begin{subfigure}[b]{0.48\textwidth}
        \centering
        \includegraphics[width=\linewidth]{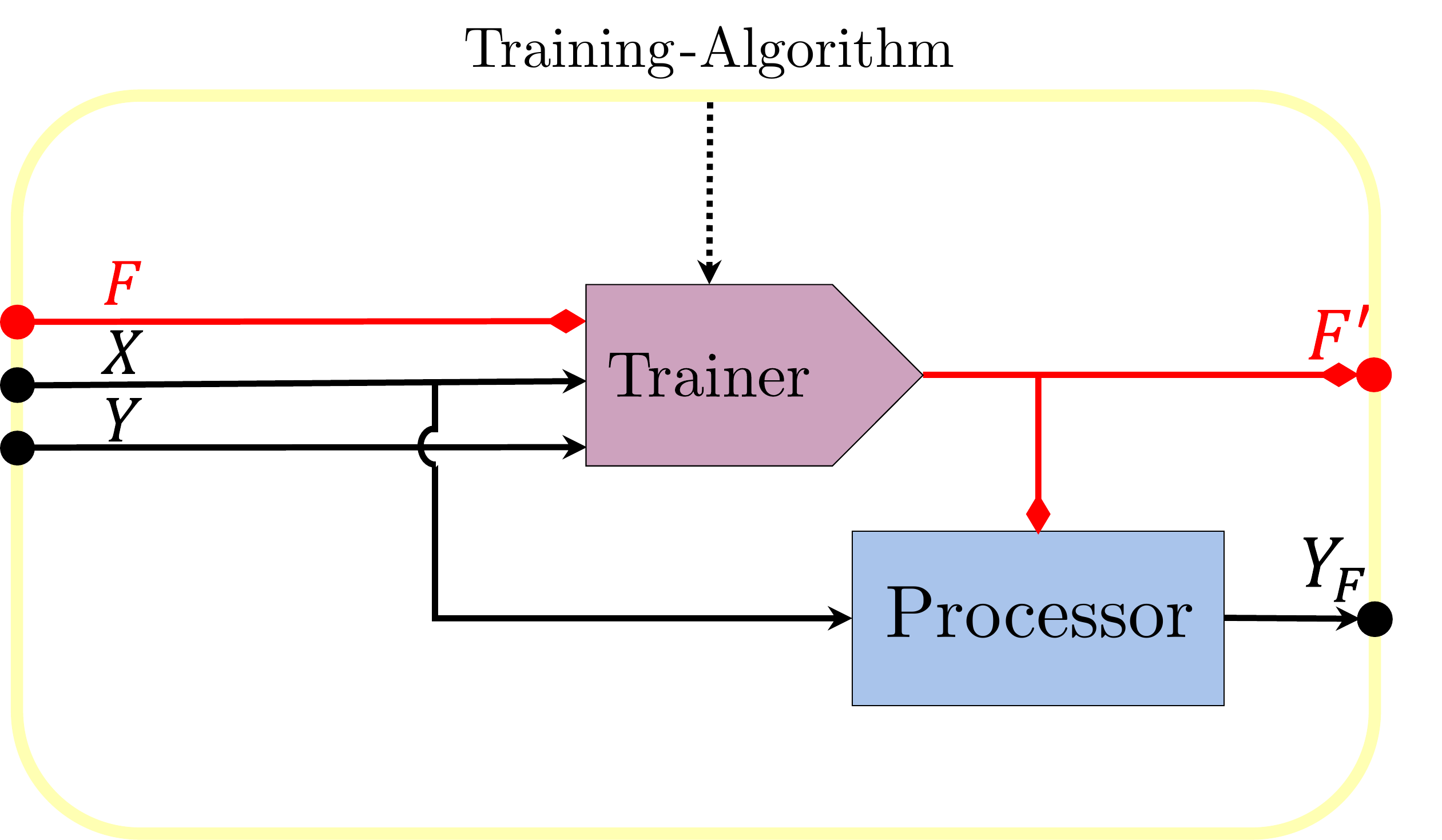}
        \caption{Non-inductive module $\FDpipe_3$}
        \label{fig:model-comb-train-pred-mod}
    \end{subfigure}
    \vfill
    \begin{subfigure}[b]{\textwidth}
        \centering
        \includegraphics[width=\linewidth]{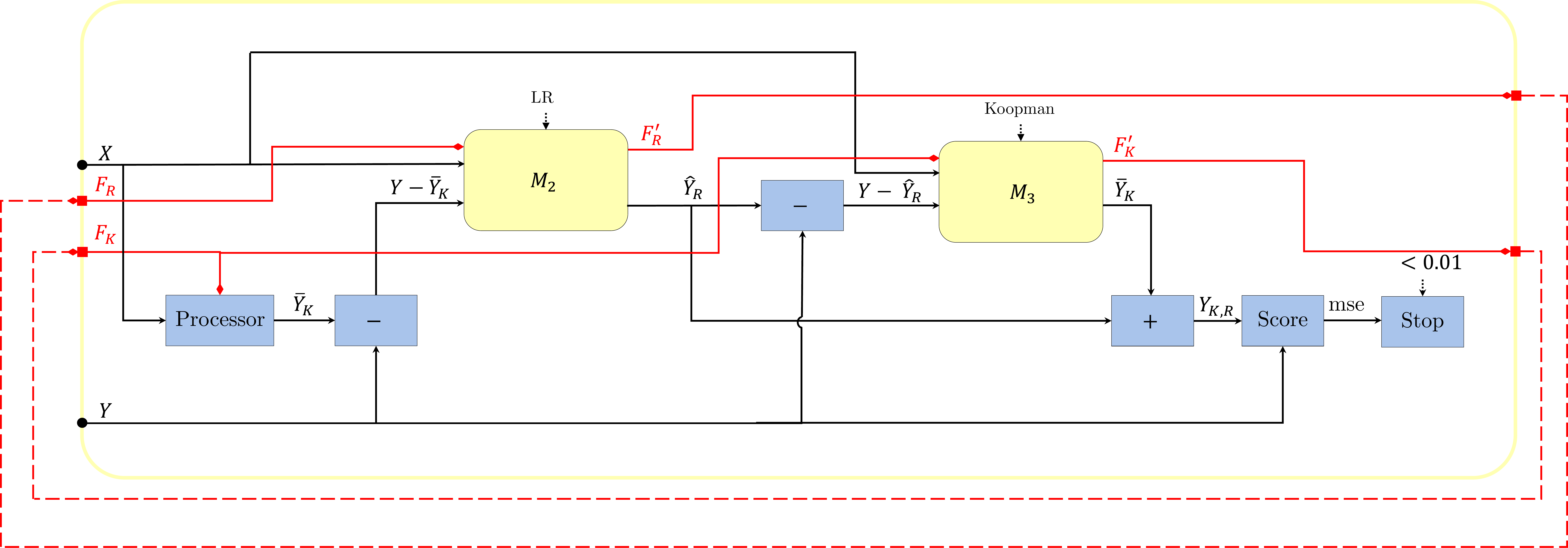}
        \caption{Inner subpipeline $\FDpipe_2$, where $\hier(\FDmod_2)=\hier(\FDmod_3)=\FDpipe_3$. $\feedback$ is represented as a dashed line connecting the corresponding recursive function ports, denoted by red squares ($\textcolor{red}{\blacksquare}$); all other ports are denoted by circles ($\bullet$)}
        \label{fig:model-comb-inner-pipe}
    \end{subfigure}
    \caption{Dual Training pipeline}
    \label{fig:model-comb-all}
\end{figure*}

We now consider a practical application of H-FDF in a complex case study. For that purpose, we illustrate how H-FDF can be used to implement and generalize a method to learn dynamical systems using "model combination"~\cite{wuNonintrusiveModelCombination2024}, that dual trains two models in a co-inductive way. 
This allows for learning faster and higher-quality models using a residual learning approach.
While standard residual learning~\cite{chinestaVirtualDigitalHybrid2020} is effective in many cases, it may converge slowly and can be suboptimal on high-dimensional systems~\cite{wuNonintrusiveModelCombination2024}.

Let $X \subset \mathbb{R}^K$ denote the state space of a dynamical system (vectors of $K$ real numbers), with states 
$\{x_k\}_{k=0}^{K-1}$ and next-state targets $\{y_k\}_{k=1}^{K}$ such that $y_k = x_{k+1} = F(x_k)$. The goal is to learn $\tilde{F}$, an approximation of $F$, by solving
\[
\tilde{F} = \arg\min_{\hat{F}\in \mathcal{G}} \mathbb{E}[L(F(x), \hat{F}(x))],
\]
where $\mathcal{G}$ is the hypothesis space of candidate models (e.g., a neural network architecture) and $L$ is the loss function measuring the distance between the predicted state $\hat{F}(x)$ and the true state $F(x)$.
In traditional residual learning, one typically learns:
\[
\tilde{F} \approx F_{R}(X) + F_{K}(X, Y - F_R(X)).
\]
By contrast, model combination iteratively refines two predictors, generating new predictions:
\[
\hat{Y}_R \gets F_R(X, Y-\bar{Y}_K), \qquad
\bar{Y}_K \gets F_K(X, Y-\hat{Y}_R).
\]
The learning process repeats until reaching a convergence criterion. Unlike standard residual learning, which performs this correction only once, model combination alternates the two updates. The final prediction adds both outputs: 
\[
Y_{K,R} = F_R(X) + F_K(X).
\]

We illustrate in Fig.~\ref{fig:model-comb-all} the implementation of this method in H-FDF. 
The topmost pipeline $P_1$ is given in Fig.~\ref{fig:model-comb-overview}, where model $F_K$ is initialized using the training dataset $(X, Y)$. 
The other model $F_R$ is initialized randomly (using the $\bot$ function). The dual training subpipeline $P_2$ is called from this initialization and dataset $(X, Y)$ to perform the core learning.

To facilitate the specification of the model combination, we use a non-inductive subpipeline $P_3$, called in \Module $\FDmod_2$ and $\FDmod_3$, to both learn a model {\em and} predict the output from this model (Fig.~\ref{fig:model-comb-train-pred-mod}).
This subpipeline uses a \Trainer algorithm (e.g., deep-learning) initialized with model $F$. After the training is performed,  the learnt model $F'$ is returned. The prediction $Y$ that results from using $F'$ on input $X$ is also returned.

Lastly, the implementation of the core dual training method is given in subpipeline $P_2$, see Fig.~\ref{fig:model-comb-inner-pipe}. It consists of training two different operators, namely (a linear regression) $F_R'$ and (a Koopman operator) $F_K'$, based on the $(X, Y)$ dynamical system dataset.     
$P_2$ is inductive: It co-trains $F'_R$ and $F'_K$, from functions $F_R$ and $F_K$ provided by the previous round (or higher level if first step). The $\feedback$ flow is denoted by dashed lines, to depict that the new models $F_R'$ and $F_K'$ are passed to be used in the next iteration. On the contrary, $X$ and $Y$ are fixed, always provided by the higher level calling the subpipeline (no feedback loop there). $P_2$ proceeds as follows:

\begin{figure*}[t!]
    \centering
    \begin{subfigure}[b]{0.95\textwidth}
        \centering
        \includegraphics[width=\textwidth]{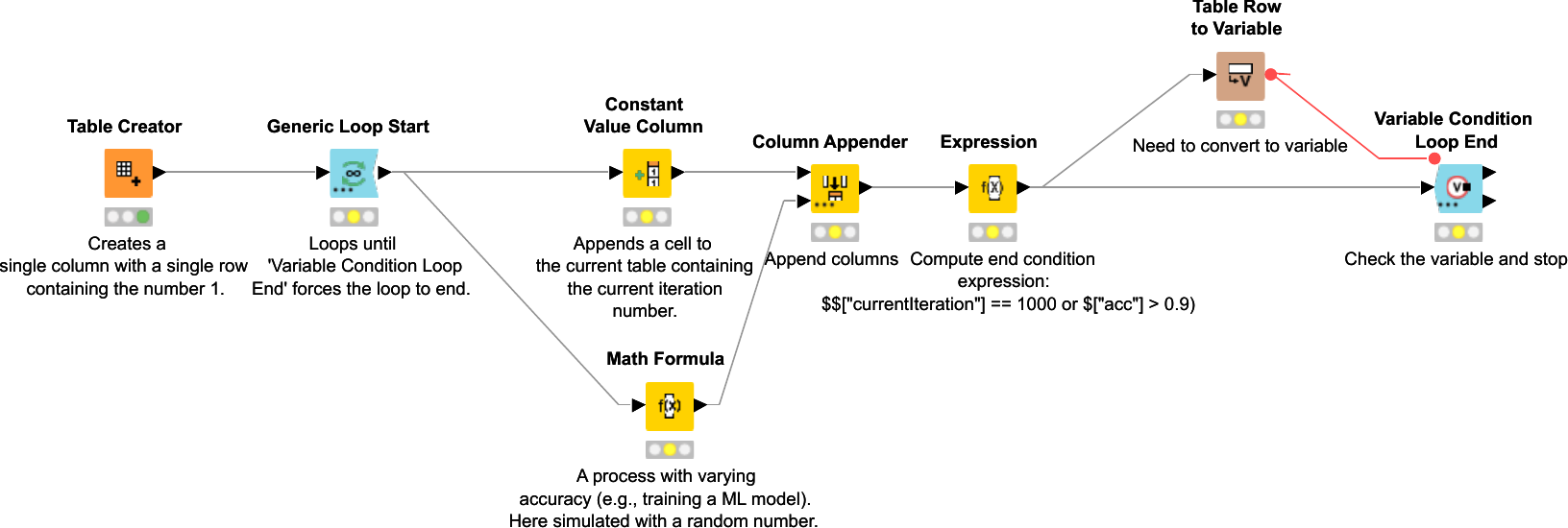}
        \caption{A KNIME pipeline to loop for at most 1000 times, checking if the accuracy (\texttt{acc}) is at least 0.9}
        \label{fig:var-cond-loop-knime}
    \end{subfigure}
    
    \vspace{1em}
    
    \begin{subfigure}[b]{0.48\textwidth}
        \centering
        \includegraphics[width=\textwidth]{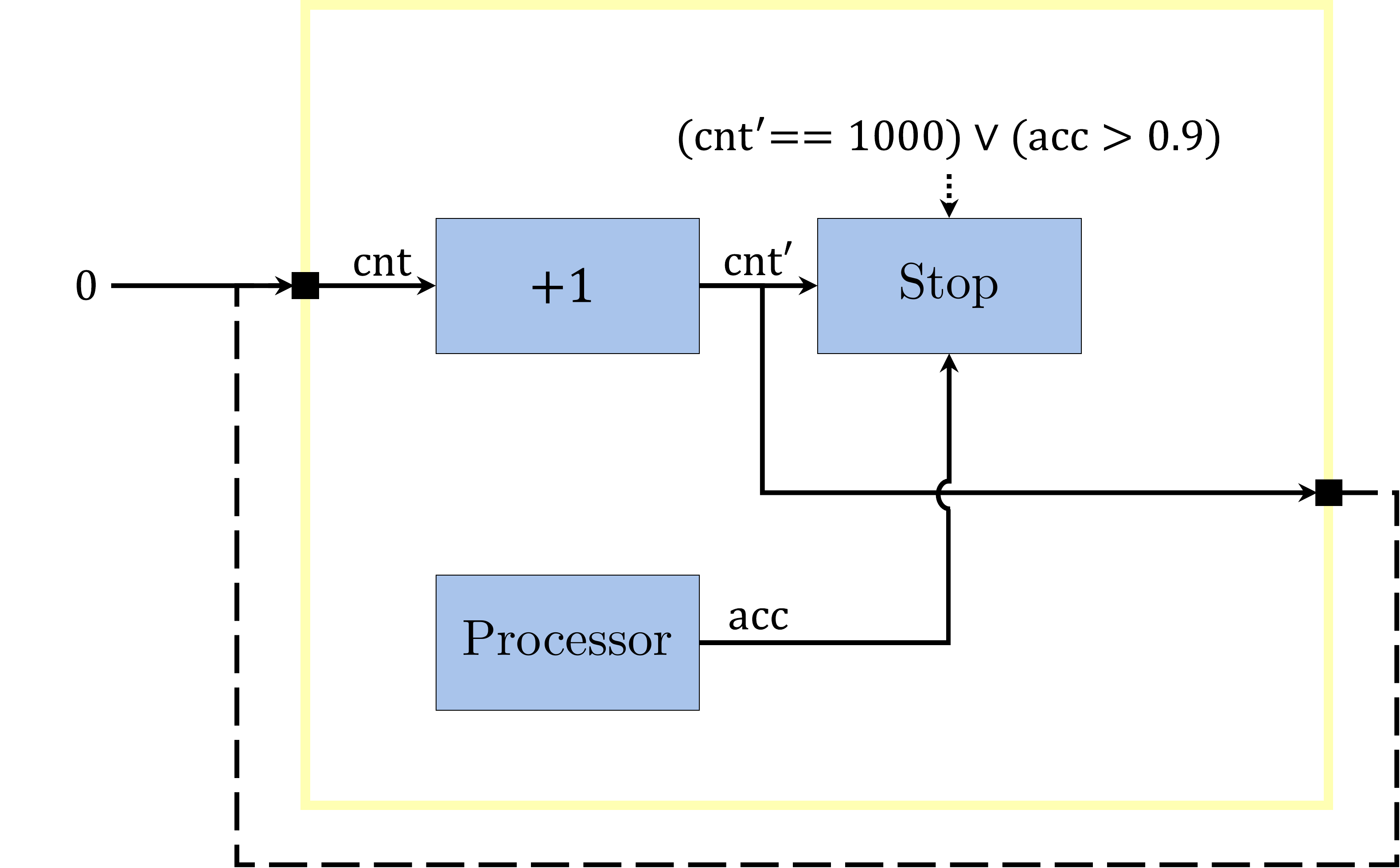}
        \caption{Equivalent H-FDF pipeline. Observe that the H-FDF pipeline is much simpler, and the stop condition is explicitly represented}
        \label{fig:var-cond-loop-fdf}
    \end{subfigure}

    \caption{Comparison of FDF pipeline in {\Builder} (top) and equivalent KNIME pipeline (bottom)}
    \label{fig:var-cond-loop}
\end{figure*}

\begin{itemize}
    \item First, $X$ is processed by $F_K$, producting the prediction $\bar{Y}_K$.
    Then $F'_R$ is trained to predict from $X$ the residual $(Y - \bar{Y}_K$), through \Module $\FDmod_2$ calling subpipeline $P_3$.
    \Module $\FDmod_2$ also outputs the prediction $\hat{Y}_R$ of the difference. 

    \item These results are used in the right side of $P_2$ (Fig.~\ref{fig:model-comb-inner-pipe}): first, the difference between $Y$ and the residual prediction $\hat{Y}_R$ is computed, and $F'_K$ is trained in \Module $\FDmod_3$ 
    to predict $Y-\hat{Y}_R$ from $X$, initialized with $F_K$. \Module $\FDmod_3$ also outputs the prediction $\bar{Y}_K$ by $F'_K$. The final prediction is the $Y_{K,R}=\hat{Y}_R + \bar{Y}_K$. 
    
    \item An MSE test ($\text{mse} < 0.01$) is performed between the predicted $Y_{K,R}$ and the expected $Y$ as a $stop$ping criteria.
    \begin{itemize}
        \item If the condition is true, iteration stops and $P_2$ returns the final model: $F_R'$, $F_K'$. 
        \item Otherwise, the subpipeline is called for another cycle. As indicated by the broken line depicting $\feedback$, the models $F_R'$ and $F_K'$ generated at this round are passed as input to the next round.
    \end{itemize}
\end{itemize}

\subsection{Comparison with KNIME}
\label{sec:loops-knime-vs-fdf}

We now compare the way to specify inductive behaviors in KNIME vs. H-FDF. 

\subsubsection{KNIME Loops} 
In KNIME, loops are declared using a matching pair of loop start and loop end nodes~\cite{bertholdKNIMEKonstanzInformation2009}. 
As stated in the KNIME documentation\footurl{https://docs.knime.com/ap/latest/analytics_platform_flow_control_guide}, the "Loop Start" node \emph{typically} defines when the iteration begins or ends, while "Loop End" specifies how the data is processed or aggregated after each iteration.
For instance, a "Counting Loop Start" allows looping for a fixed number of times, and it can be paired, among others, with "Loop End" to append incoming tables row-wise or "Loop End (column append)" to append column-wise. 
However, some "Loop End" nodes, such as "Variable Condition Loop End", can also terminate the loop based on a specific condition, which must be represented by a KNIME flow variable (similar to parameters in H-FDF). 

KNIME's approach results in a large catalog of loop nodes, each with specific use cases. The user must also carefully pair loop start and loop end nodes (e.g., a Recursive Loop Start must always be paired with a Recursive Loop End).

\subsubsection{Comparative Analysis}
We compare the KNIME and H-FDF approaches in Fig.~\ref{fig:var-cond-loop}, which shows a pipeline where the loop stops when either a process accuracy (simulated via a random number generator) reaches at least 0.9, or the loop repeats for 1000 iterations.

Observe that KNIME's loop creates a \emph{coupling between data flow and control flow} in the loop definition: loop start and loop end nodes jointly determine how data is passed between iterations (data flow) and how many times to repeat the loop (control flow). 
Importantly, in KNIME, there is no generic function flow between rounds. In some cases, like active learning, specific looping boxes exist. Otherwise, the functions must first be converted into data, passed, and converted back as functions. 
By comparison, in H-FDF, data and functions passed from one round to the next are explicitly and graphically represented by using $\feedback$ flow.

This coupling in KNIME necessitates extensive rewiring for even minor logic changes. While a pipeline to loop for 1000 iterations can be easily implemented in KNIME using a "Counting Loop Start" paired with a "Loop End", implementing a more complex pipeline, which loops for 1000 times or until accuracy reaches 0.9, requires (Fig.~\ref{fig:var-cond-loop-knime}\footnote{Available as \emph{knime-vs-fdf/iter-var-knime.knwf} in the paper artifact.}): 
\begin{itemize}
    \item swapping the "Counting Loop Start" for a "Generic Loop Start",
    \item explicitly computing the end condition using both the current iteration and process accuracy,
    \item casting the result to a flow variable for use in the "Variable Condition Loop End", and 
    \item replacing the "Loop End" with a "Variable Condition Loop End".
\end{itemize}

By contrast, H-FDF separates these concerns by introducing an explicit stop port and requiring it to be explicitly computed and represented in the graph (Fig.~\ref{fig:var-cond-loop-fdf}\footnote{Available as \emph{knime-vs-fdf/iter-var-hfdf.svg} in the paper artifact.}). Therefore, once a new process is inserted in the pipeline, it suffices to adapt the stop condition computation to consider both the accuracy and the loop counter.

\section{Conclusion} \label{sec:conclusion}

This paper presents a user evaluation of the FDF framework and its implementation, {\Builder}, alongside an FDF extension to support iterative pipelines: H-FDF (Hierarchical Function+Data Flow).The user evaluation assesses the perceived usability and the feature adequacy of the current framework, identifying its strengths and key improvement areas to inform future development.
To address the lack of explicit iteration in FDF, H-FDF introduces a specialized \Module box. This component supports iteration until a stop condition occurs, while ensuring the global pipeline remains acyclic and easy to reason about.

The user study indicates that participants perceive FDF and {\Builder} as an intuitive, easy-to-use tool. This is quantified by a \emph{good} usability rating, with a median SUS score of 72.5 (mean: 68.9 $\pm$ 15.2). 
Qualitative feedback highlights that {\Builder} is an efficient, integrated platform for rapid experimentation with an easy-to-use "plug-and-play" graph interface.
Improvement areas include adding tooltips and integrated documentation.

Regarding H-FDF, we formalize its syntax and semantics and illustrate its utility through a complex dual-training case study. This application shows that H-FDF can flexibly combine models co-inductively. Furthermore, a comparative analysis with KNIME suggests that H-FDF's explicit loop conditions reduce the required effort to evolve the iteration strategy.

Future work focuses on the full implementation of H-FDF within the {\Builder} modeling environment. We also aim to enhance user experience by integrating contextual documentation and supporting real-time pipeline inspection during execution.

\backmatter

\bmhead{Code availability} 
To improve reproducibility, supplementary material is available on Zenodo: \url{https://doi.org/10.5281/zenodo.21154924}. This includes the user study instruments, results, and analysis scripts. {\Builder} source code can be found at \url{https://github.com/CPS-research-group/descartes-builder}. We evaluated the release v0.2 in the user study with the corresponding installer at \url{https://doi.org/10.5281/zenodo.18757324}.

\bmhead{Acknowledgements}
We would like to thank Amine Ammar (ENSAM), CETIM, and Joël Mouterde (SKF) for providing the DT prototypes.
This research is part of the program DesCartes and is supported by the National Research Foundation, Prime Minister's Office, Singapore under its Campus for Research Excellence and Technological Enterprise (CREATE) program.

\bibliography{cite,fdf,sens}

\newpage

\begin{appendices}

\section{Evaluation Material}

\begin{table*}[t!]
    \centering
    \caption{Complete questionnaire for {\Builder} usability (based on the SUS questionnaire~\cite{brookeSUSQuickDirty1996})}
    \label{tab:sus_builder_full}
    
\begin{tabularx}{\textwidth}{llX}
    \toprule
    ID  & Short label (positive only) & Full SUS item (positive/negative version)                                                 \\ \midrule
    $Q_1$  & Intended frequency of use   & I think that I will use {\Builder} rather than other methods.                             \\
    $Q_2$  & System simplicity           & I found {\Builder} unnecessarily complex.                                                 \\
    $Q_3$  & Ease of use                 & I thought {\Builder} was easy to use.                                                     \\
    $Q_4$  & User independence           & I think that I would need the support of a technical person to be able to use {\Builder}. \\
    $Q_5$  & Functions well integrated   & I found the various functions in {\Builder} were well integrated.                         \\
    $Q_6$  & Design consistency          & I thought there was too much inconsistency in {\Builder}.                                 \\
    $Q_7$  & Learnability                & I would imagine that most people would learn to use {\Builder} very quickly.              \\
    $Q_8$  & Navigation smoothness       & I found {\Builder} very cumbersome to use.                                                \\
    $Q_9$  & User confidence             & I felt very confident using {\Builder}.                                                   \\
    $Q_{10}$ & Low learning effort         & I needed to learn a lot of things before I could get going with {\Builder}.    \\ \bottomrule
\end{tabularx}
 \end{table*}

\subsection{Informed Consent} \label{app:informed-consent}

Participants were requested to sign an informed consent document that covers the basic ethical aspects considered by the project.

\emph{
You are invited to participate in a research study evaluating {\Builder}, a tool for developing machine-learning pipelines for digital twins. We aim to test the hypothesis that the abstraction and tooling proposed can facilitate the development of machine learning-based digital twins.
The experiment will require about 2 hours of your time and involves a short tutorial, a hands-on task using the tool, and a survey about your experience. All the collected data will be anonymized and kept confidential. Your feedback will help improve the tool and shape its future features. 
Note that your participation is entirely voluntary; you will not receive compensation for your time, and you can withdraw from the study at any time without penalty. 
If you agree, your anonymized data may be stored for future research, but you can opt out at any time.
For questions or concerns, you can contact the study team or the NTU Institutional Review Board. By signing this consent form, you are giving your \emph{informed and voluntary consent} to take part in this research study, and you confirm you understand the study's purpose, procedures, and your rights.}

\subsection{User Survey} \label{app:user-survey}

\subsubsection{Background}

\begin{enumerate}
    \item \textbf{What is your current role?}
    \begin{itemize}
        \item Researcher
        \item Student
        \item Engineer
        \item Industry Practitioner
        \item Other
    \end{itemize}
    
    \item \textbf{What is your primary area of expertise?}
    \begin{itemize}
        \item Artificial Intelligence
        \item Systems Engineering
        \item Software Engineering
        \item Digital Twins
        \item Other
    \end{itemize}
    
    \item \textbf{Please rate your familiarity with the following concepts and tools (Rating: 1 = None to 5 = Expert):}
    \begin{itemize}
        \item Machine Learning (ML)
        \item Digital Twins
        \item Reduced-order modeling
        \item System modeling tools (e.g., Simulink, Modelica)
        \item Engineering simulation tools (e.g., Ansys, OpenFOAM)
    \end{itemize}
\end{enumerate}

\subsubsection{Tool Usability}

\begin{enumerate}[resume]
 \item \textbf{Please indicate your level of agreement with the statements in Table~\ref{tab:sus_builder_full} (Rating: 1 = Strongly Disagree to 5 = Strongly Agree).}
\end{enumerate}

\subsubsection{Experience with {\Builder}}
\begin{enumerate}[resume]
    \item \textbf{How user-friendly did you find the interface for the following tasks? (Rating: 1 = Very Demanding to 5 = Very Intuitive):}
    \begin{itemize}
        \item Box Parametrization
        \item Pipeline Modeling
        \item Result Visualization
        \item Warning Messages
    \end{itemize}
    
    \item \textbf{Which challenges did you encounter while using the tool?}
\end{enumerate}

\subsubsection{Open Feedback}
\begin{enumerate}[resume]
    \item \textbf{What features or improvements would most enhance your experience with DesCartes Builder?}
    \item \textbf{Any additional comments or suggestions regarding usability, ML integration, or digital twin functionality?}
\end{enumerate}
 \section{Sensitivity Analysis} \label{sec:sensitivity-analysis}

\SetKwInput{KwParam}{Parameter}
\SetKwData{Out}{Out}
\begin{algorithm*}[t!]
    \DontPrintSemicolon
    \caption{Feature-Based Sensitivity Analysis for Model Robustness}
    \label{alg:sens-analysis}
    \KwIn{Trained model $f$}
    \KwParam{
        $M$, resolution for querying $f$; 
        $N$, number of samples for sensitivity analysis; 
        $tgt$, target $f$ output feature: $f(x)[tgt]$; 
        $[low, high]$, bounds of $f$ input domain (used to define $\mathcal{X}$);
        $div$, divisor to define local neighborhood space.
    }
    \KwOut{$N$ input pairs $(x_1, x_1'),\ldots, (x_N, x_N')$ and corresponding output pairs $(f(x_1), f(x_1')),\ldots,(f(x_N), f(x_N'))$ such that $\Delta(f(x_i), f(x_i'))$ is maximized.}
    \SetKwData{rndFull}{rndFull}
    \SetKwData{attrFull}{attrFull}
    \SetKwData{topFull}{topFull}
    \SetKwData{rndDiv}{rndDiv}
    \SetKwData{query}{query}
    \SetKwData{attrDiv}{attrDiv}
    \SetKwData{topDiv}{topDiv}
    \SetKwData{diff}{diff}
    \SetKwData{grid}{grid}

    \tcp{Identify most influential samples globally}
    \rndFull $\gets$ $M$ random samples uniformly at random in range $[low, high]$\;
    \attrFull $\gets$ attribution scores for the feature $f(x)[tgt]$ over \rndFull\;
    \topFull $\gets$ top-$N$ samples with highest attribution from \attrFull\;
    \tcp{Refine around \topFull influencial regions}
    \ForEach{$s \in \topFull$}{
        \query $\gets$ neighborhood around $s$ of size $(high - low) / div$\;
        \rndDiv $\gets$ new $M$ random samples within \query\;
        \attrDiv $\gets$ attribution score for $f(x)[tgt]$ over \rndDiv\;
        \topDiv $\gets$ top-$N$ samples with high attribution score from \attrDiv\;

        \tcp{Finer sampling in a grid around \topDiv}
        \ForEach{$s' \in \topDiv$}{ $j_1, j_2 \gets$ top-2 most influencial input features\; \grid $\gets M \times M$ grid varying $x[j_1], x[j_2]$ over $[low, high]$ and fixing other dimensions\;
            $\Out \gets \{ f(x) \mid x \in \grid \}$ \tcp*[h]{evaluate $f$ on all $x \in \grid$}\; 
            Define the neighborhood $\mathcal{N}(\cdot)$ as the set of grid-adjacent elements for $x \in \grid$ and $y \in \Out$\;
            \tcp{Maximize local output discrepancy}
            $\diff \gets \argmax_{y \in \Out, y' \in \mathcal{N}(y)} \| y - y' \|_\infty$\;
            \tcp{Retrieve corresponding neighboring input-output pairs}
            $x_i \gets \grid[\diff]; x_i' \gets \mathcal{N}(x_i)$ \;
            $y_i \gets \Out[\diff]; y_i' \gets \mathcal{N}(y_i)$ \;
 }
 }
\end{algorithm*}

The standard way of testing ML-based DTs is to assess the model's quality metric on the hold-out test data. The most common metric is the mean-squared error (MSE) between the ML model prediction and the test data~\cite{raiDrivenDataDerived2020}. 
However, for ML models, particularly neural networks (NN), \emph{current validation practices may not be sufficient}, since they explore the PT input space using random sampling or predefined grids. Indeed, a highly accurate model (e.g., low MSE) may still have regions with large discrepancies between the DT and the PT\@. This issue is due to at least two key factors. 
Firstly, NNs are known to be brittle to \emph{adversarial attacks}, where small, often imperceptible perturbations (e.g., altering a single pixel) can result in significant changes in the output~\cite{szegedyIntriguingPropertiesNeural2014,suOnePixelAttack2019}.
Secondly, the design of experiments (DoE) used for surrogate design may not adequately cover critical regions of the input space~\cite{garudDesignComputerExperiments2017}. For instance, consider the DT prototype for material strain prediction (Section~\ref{sec:dtp-material-strain}). When dealing with aluminum, which exhibits a highly nonlinear behavior under cyclic loading~\cite{pisapiaExperimentalCampaignStructural2023}, an inadequate DoE might overlook these regions, leading to inaccurate predictions. 

To mitigate this, we aim to identify \emph{regions where the surrogate does not faithfully represent the physics model}. The naive way to address this problem would be to perform numerous physics simulations, comparing them with the ML surrogate directly. However, this is impractical due to the high computational cost of the physics model. 
Hence, our strategy is to first \emph{analyze the surrogate thoroughly}, identifying regions with low robustness, i.e., regions where small perturbations in input yield large changes in output.
These regions are expected to be meaningful candidates for prompting the physics model since most data sources and physical models are smooth and bounded, allowing the construction of a compact local representation~\cite{burgessUnderstandingDisentangling$beta$VAE2018, chenNeighborhoodGeometricStructurePreserving2022}. If we can identify these regions, we could, for instance, include more points in the DoE for them and \emph{correct} the model by placing more emphasis on these critical regions during training.

Formally, let $f:\mathcal{X} \to \mathbb{R}^d$ denote the ML surrogate, where $\mathcal{X}$ is the input space of the ML surrogate and $\mathbb{R}^d$ is the $d$-dimensional output space. For a given input $x \in \mathcal{X}$, the goal of sensitivity analysis is to identify a neighboring point $x'$ such that the output difference $\Delta(f(x), f(x'))$ is maximized (e.g., $\ell^\infty$-norm, commonly used in adversarial analysis~\cite{madryDeepLearningModels2019}):
$$
x,x' =  \argmax_{x,x'\in\mathcal{X}} \Delta(f(x), f(x')), 
 \text{where} \|x-x'\| < \varepsilon
$$

Since exhaustive search is infeasible, we propose a scalable heuristic (Algorithm~\ref{alg:sens-analysis}) that uses feature attribution methods~\cite{sundararajanAxiomaticAttributionDeep2017,lundbergUnifiedApproachInterpreting2017} to search regions of high output sensitivity. This method yields candidate point pairs that expose brittle regions in the model $f$.

The algorithm consists of three key steps: 
\begin{enumerate}
    \item \textbf{Lines 1--3:} Sample uniformly at random the full input space of $f$ and obtain feature attribution scores using, e.g., the Integrated Gradients method~\cite{sundararajanAxiomaticAttributionDeep2017}. This will lead to an importance score for the input features $\mathcal{X}$ of $f$ on the specific output target $tgt$ considered. 
    \item \textbf{Lines 4--8:} Based on the attribution score, refine the focus from the full input space $\mathcal{X}$ to highly sensitive regions. Obtain the feature attribution score as in the previous step for these sub-regions. The rationale is to assess the subset of features that are most important in this sub-region, which may be different from that of the full space. 
    \item \textbf{Lines 9--16:} On the highly sensitive regions, evaluate the model $f$ by fixing the top-2 most influential features and varying the remaining features. Identify points in the model output $\Out$ such that their difference is maximized. 
\end{enumerate}

The implementation is integrated into the {\Builder} back-end\footurl{https://github.com/eduardoconto/kedro-umbrella/blob/main/kedro_umbrella/library/sensitivity.py} and relies on the Captum~\cite{kokhlikyanCaptumUnifiedGeneric2020} library.
 \end{appendices}
 \end{document}